\newcommand{\beq}{\begin{equation}}
\newcommand{\eeq}{\end{equation}}
\newcommand{\beqau}{\begin{eqnarray*}}
\newcommand{\eeqau}{\end{eqnarray*}}
\newcommand{\beqa}{\begin{eqnarray}}
\newcommand{\eeqa}{\end{eqnarray}}
\newcommand{\fig}[1]{Figure~\ref{#1}}
\newcommand{\tab}[1]{Table~\ref{#1}}

\newcommand{\half}{\frac{1}{2}}

\newcommand{\fltfig}[3]{
\begin{minipage}{15cm}
\begin{figure}
\begin{center} 
{\large \caption{ \label{#3} #1 }} 
#2
\end{center}
\end{figure}
\end{minipage}
}

\newcommand{\flttab}[3]{
\begin{table}[htbp]
{\large \caption{ \label{#3} #1 }}
\begin{center} 
#2
\end{center}
\end{table}
}
\newcommand{\xorg}{0}
\newcommand{\yorg}{0}
\newcommand{\state}[6]{
   \put(\xorg,0){ \put(#6,-10){\makebox(0,0)[t]{${ #1}$}}               }
   \put(\xorg,\yorg){ \put(#6,#3){\circle*{10}}                         }
   \put(\xorg,\yorg){ \put(#6,#3){\line(0,1){#4}}                       }
   \put(\xorg,\yorg){ \put(#6,#3){\line(0,-1){#5}}                      }
   \put(\xorg,\yorg){ \multiput(#6,#2)(20,0){2}{\line(1,0){10}}         }
   \put(\xorg,\yorg){ \multiput(#6,#2)(-20,0){3}{\line(1,0){10}}        }
                     }
\newcommand{\spectrum}[7]{
   \put(\xorg,0){ \put(0,0){\line(0,1){#1}}                             }
   \put(\xorg,\yorg){ \multiput(-20,#2)(0,500){2}{\line(1,0){40}}       }
   \put(\xorg,0){ \multiput(-10,0)(0,100){8}{\line(1,0){20}}            }
   \put(\xorg,\yorg){ \put(-50,#2){\makebox(0,0)[r]{#3}}                }
   \put(\xorg,\yorg){ \put(-50,#4){\makebox(0,0)[r]{#5}}                }
   \put(\xorg,\yorg){ \put(-50,#6){\makebox(0,0)[r]{#7}}                }
 
                        }
\newcommand{\epsfaxhax}[2]{
        \centerline{
          \hspace{-20pt}
          \epsfxsize=225pt
          {\epsfbox{#1}}
          \hspace{-15pt}
          \epsfxsize=225pt
          {\epsfbox{#2}}}
}
\newcommand{\epsfaxhaxhax}[3]{
        \centerline{
          \hspace{-20pt}
          \epsfxsize=160pt
          {\epsfbox{#1}}
          \hspace{-15pt}
          \epsfxsize=160pt
          {\epsfbox{#2}}
          \hspace{-10pt}
          \epsfxsize=160pt
          {\epsfbox{#3}}}
}
\newcommand{\pspicture}[1]{\centerline{\setlength\epsfxsize{225pt}\epsfbox{#1}}}

\documentstyle[aps,preprint,tighten,epsf]{revtex}
\begin{document}
\preprint{
\begin{minipage}{4.5cm}Edinburgh Preprint 99/2\\
GUTPA-99-02-05\\\end{minipage}}

\draft
\author{Peter Boyle
\thanks{Present Address: Department of Physics and Astronomy, 
University of Glasgow, Glasgow}}
\address{Department of Physics and Astronomy, University of Edinburgh, Edinburgh, UK}

\author{{\bf UKQCD Collaboration}}
\date{\today}
\title{The heavy quarkonium spectrum from quenched lattice QCD}
\maketitle
\begin{abstract}
We present results of simulations of the quenched quarkonium spectrum 
at two values of the lattice spacing and for quark masses
around $m_c$ using the tadpole 
improved clover action. Attention is focussed
on the lowest lying S and P states, 
and the triplet fine structure is
obtained for the first time using a relativistic action.

\end{abstract}

\section{Introduction}
Lattice QCD simulations provide a systematically improvable
method for predicting low energy phenomenology directly from QCD,
providing a means to verify that QCD correctly reproduces
experiment.
We are primarily interested here in the simulation of 
charmonium, which remains a challenge for lattice QCD since the 
quark mass is sufficiently large to be difficult to resolve
on coarse lattices, while the system is rather relativistic, 
causing problems for the NRQCD approach.

The angular momentum states calculated here are the $^1S_0$, $^3S_1$, $^3P_0$, $^3P_1$, and $^3P_2$
of lowest radial excitation. This is more extensive than typical lattice
calculations using a relativistic action,
 which have generally been limited to the $^1S_0$, $^3S_1$ and spin-averaged S-P splittings 
\cite{kronfeld_mflgt,UKQCDhyperfine,sara_M1_M2_discrepancy,sara_orbital}, 
though Bhattacharya et al \cite{rajan} produce a value for the
$\chi_{c1} - \chi_{c0}$ fine splitting using unimproved Wilson quarks.

\subsection{Simulation Techniques for  Charm and Beauty}

A discrete action approximating the continuum action is used, 
where finite differences are substituted for derivatives, 
summations for integrals and so on, giving the $O(a)$ improved
Wilson quark action \cite{SW,heatlie}
\beq
S_{\rm latt} = \frac{1}{2\kappa} \sum\limits_{x,y} \bar{\psi}(x) M(x,y) \psi(y),
\eeq
\beq
M(x,y) = \left[1 - \frac{i \kappa C_{SW}ag}{2} \sigma_{\mu\nu} F_{\mu\nu}\right] \delta_{x,y} - \kappa \sum\limits_{\mu=1}^4
U_\mu(x)(1-\gamma_\mu)\delta_{y,x+\mu} + U^\dagger_\mu(x-\mu)(1+\gamma_\mu)\delta_{y,x-\mu},
\eeq
where $\kappa = \frac{1}{2 m_0 a + 8}$, and the gauge group 
elements $U_\mu(x) = e^{iagA_\mu(x+\half\hat{\mu})}$.
An expansion of the discrete action in terms of continuum fields and their
derivatives gives a natural classification
of errors in powers of the lattice spacing $a$
\beq
S_{\rm latt}\left[\bar{\psi}(x),\psi(x),U_\mu(x)\right]
 = S_{\rm QCD}\left[\bar{\psi}(x),\psi(x),A_\mu(x)\right] + a S_1\left[\bar{\psi}(x),\psi(x),A_\mu(x)\right] + O(a^2).
\eeq
We have chosen a lattice action which reduces to QCD as $a\rightarrow 0$,
and may eliminate the leading errors due to $S_1$
by adding appropriate (higher dimension) operators.
The clover term 
\beq
- \bar{\psi} C_{SW} \frac{iag}{4} \sigma_{\mu\nu} F_{\mu\nu} \psi.
\eeq
gives an $O(a)$ contribution to the action so is irrelevant to the 
continuum limit; its coefficient may be selected
to minimise the leading discretisation effects contained in $S_1$.
The fixing of the clover coefficient has been performed in various ways -
to different orders in perturbation theory \cite{SW,heatlie,CSW_1loop},
using mean-field improvement \cite{tadpole}, and most recently using non-perturbative
improvement criteria \cite{luschernp}, giving a much improved continuum scaling behaviour
of the light hadron sector \cite{hartmut_lat97}. 
This categorisation of errors in powers of 
the lattice spacing $a$, however, introduces an upper limit
for quark masses that can be accurately simulated, $a m_Q \ll O(1)$.
Within the quenched approximation contemporary simulations typically
have $am_{\rm charm}\ge \frac{1}{2}$, 
leaving them apparently susceptible to significant 
discretisation effects, which can only be removed
by brute force extrapolation to the continuum limit. 
In recent years the problem of simulating heavy quarks in lattice QCD
has therefore been tackled by discretising the effective theory, NRQCD --
essentially performing a double (power counting) expansion in both the lattice spacing
and in $\frac{v^2}{c^2}$ \cite{LepagePowerCounting}, 
producing very wide ranging spectrum predictions for
quarkonia \cite{cdavies_upsilon_long,ajl_cc,davies_bc} . 
However, the non-relativistic approach has recently been demonstrated to suffer
from significant relativistic corrections for charmonium \cite{Trottier}, leading
to inconsistent results when comparing different orders of the nonrelativistic
action \cite{ChristineTsukuba}, while the relativistic 
corrections to $\Upsilon$ appear to be well under control. 

If we make a power counting expansion, in the manner of Fermilab 
\cite{kronfeld_mflgt}, of the improved Wilson action in both $v^2$ and 
$a$ simultaneously, however, we can
see that the situation for heavy valence quarks is not as bad 
as na\"ively supposed. 
Consider the tree-level low momentum expansion of the energy,
(while this applies to the free case, the errors at large $m_0 a$ are certainly
indicative of 
errors in the interacting theory), 
\beq
E(p) = m_1 + \frac{p^2}{2 m_2} - \frac{i\sigma\cdot{\bf B}}{2m_B} + \frac{p^4}{8 m_4^3} +
A^{RS} + B^{SD} + \ldots
\eeq
where
\beq
a m_1  = \log{(1+ a m_0)},
\eeq
\beq
\label{WilsonDispersion}
a m_2 =  e^{a m_1} \frac{\sinh (a m_1)}{1+\sinh (a m_1)}
\eeq
\beq
a m_B = e^{a m_1} \frac{\sinh (a m_1)} {1+C_{SW}\sinh (a m_1)}
\eeq
\beq 
a m_4 = e^{a m_1} \frac{\sinh (a m_1) }{\left[ 1 + e^{a m_1} \sinh (a  m_1) 
( 1 + \sinh^2 (a m_1) ) + 2 e^{2 a m_1} \sinh (a m_1) \right] ^ {\frac{1}{3}} }
\eeq
where the $A^{RS}$ term violates rotational symmetry at $O(\frac{1}{m^3})$ and 
$B^{SD}$ contains 
order $v^6$ spin-dependent corrections.
 
In this light, the accuracy of the action depends on both the
size of $a m$ and on the size of $v^2$. In a sufficiently non-relativistic
system the above Hamiltonian will, at tree level, accurately describe a particle
of mass $m_2 = m_B$. If $a m \ll 1$ the action is approximately Lorentz
symmetric \footnote{assuming that $a \Lambda_{QCD} \ll 1$} 
and the coefficients of higher order terms in the above non-relativistic 
action will accurately match those of the continuum expansion. 
The essential point for charm however is that
we have to estimate the size of errors in the subleading terms relative
the terms containing $m_2$ and $m_B$. The errors will depend on both the
difference between the higher order coefficient from that of the continuum
Hamiltonian for a particle of mass $m_2$, and on the size of 
an $O(v^{2n})$ prefactor. At tree level, therefore, we can estimate the 
fractional error introduced in the kinetic energy due to discretisation effects in
the action as follows.

\beq
\Delta_{\rm kin} = \frac 
{\frac{\langle p^4 \rangle}{ 8 m_4^3 } - \frac{\langle p^4 \rangle}{ 8 m_2^3 }}
{ \frac{\langle p^2 \rangle}{2 m_2} }
\eeq

Taking a charm quark mass of around $1.5$ GeV and
an estimate for $\langle v^2 \rangle \simeq 0.3 \frac{a m_{\rm charm}}{a m}$ (which
approximates the data in \cite{BaliBoyle} with sufficient accuracy to satisfy
our needs), we plot the fractional error in the kinetic energy, 
$\Delta_{\rm kin}$ as a function
of $a m$ for both of the lattice spacings we use in figure \ref{FigKineticError}.
The leading order error in the kinetic energy
is estimated to be between $5\% - 10\%$.
We note that had we used the pole mass, the leading correction
would have arisen from the $m_1 - m_2$ discrepancy rather than the
$m_2 - m_4$ discrepancy, and therefore would not have been suppressed 
by $v^2$. Both analyses would, of course, have given the same continuum limit;
however the use of $m_2$ will for this reason have a milder lattice
spacing dependence.

The disadvantage of any simulation involving Wilson quarks with $a m_0 \ge O(1)$ 
is that the  $a$ dependence  of observables is not likely to have a simple
low order polynomial form. There is no hierarchy
in the orders of $a$ due to terms $(m a)^n$, \footnote{we are dealing with on-shell correlation
functions of valence quarks} which is manifested in the non-linear dependence
of the various mass definitions on the bare quark mass. 
These different mass definitions rapidly become identical as $m a$ is reduced
below $1$, so that, for charm, studying the $a$ dependence is sensible with
current lattice spacings.
For bottomonium results, however, $a m_0 \ge 1$ in both simulations, and
we must rely on the power counting estimates of the correction
terms to quantify  errors. Continuum scaling plots provide
only circumstantial evidence that the systematic errors in the bottom simulations
are controlled. A more rigorous approach would be to form the 
union of bounds estimated for different error terms on each of the lattices; in this
respect simulations in which $a m_0 \ge 1$ must be 
treated in a similar way to effective theories --
the fact that the continuum limit exists for the Fermilab approach \cite{kronfeld_mflgt} 
is not in practice an advantage for bottom simulations since we do not
know how to extrapolate \emph{from this regime}.

Ideally we would make a corresponding estimate for the errors in the
spin-dependent pieces of the action, but the subleading spin-dependent
term for the Wilson action has not presently been calculated. However, it is
thought that a similar size systematic error is likely in the spin-dependent 
Hamiltonian.

Some progress has recently been made in treating Wilson type quarks
to one loop in perturbation theory for arbitrary quark mass.
In particular Kronfeld et al \cite{kronfeld_self_energy} have calculated the
1-loop renormalisation of the terms $m_1$ and $m_2$, showing that the discrepancy
is dominated by the tree level result.
The Kuramashi \cite{kuramashi_Zv}
calculation gave bilinear current renormalisation constants for Wilson
quarks, and somewhat interestingly, noted that it was necessary to match 
$m_2$ to the continuum pole mass in order to obtain the same IR behaviour for
the vertex and wavefunction in the lattice theory as in the continuum theory.
In order to obtain more rigorous control of the errors in charmonium simulations
it would, of course, be interesting to have a one-loop calcution
for $m_4$, and (at least) a tree level estimate of the subleading spin-dependent
pieces for Wilson type actions.

In the remainder of this paper we present a study of 
quarkonium systems using the tadpole 
improved clover action \cite{tadpole,heatlie} at both $\beta=6.0$
and $\beta=6.2$, corresponding to $a\simeq 0.1$ fm and $a \simeq 0.07$ fm, 
analysed in the manner suggested by \cite{kronfeld_mflgt}.
In the calculation at $\beta=6.0$
similar operators to those used in NRQCD calculations are applied
enabling us to produce a wider ranging spectrum than previous
relativistic calculations.
The scaling behaviour of the hyperfine splitting
between the two lattices is discussed, and estimates of the
consistency of continuum quenched results with experiment are made.
We note that while
El-Khadra, Kronfeld, Mackenzie and
others \cite{kronfeld_mflgt,kronfeld_self_energy}
propose improving the action for on shell quantities by breaking $O(4)$ rotational
symmetry in the Fermilab action, 
their simulations to date have involved the same re-interpretation of the
standard relativistic quark action as we use here, and the results
are therefore directly comparable.

\section{Quenched Quarkonium Spectrum at $\beta=6.0$}

\subsection{Simulation Details - $\beta=6.0$ calculation}
The first simulation was performed 
using 499 sample gauge configurations
from a quenched distribution at $\beta=6.0$
on a $16^3\times48$ lattice. Five heavy quark
masses were used, as given in \tab{TabHeavy60Kappas}, with
the tadpole improved clover action with $C_{SW} = 1.47852$, corresponding to 
$u_0 =  0.8778$ taken from the plaquette.

The timesliced residue
\beq
|r_t|^2 = \frac{ \sum\limits_{x\in L^3} | \sum_{x^\prime,t^\prime} M(x,t;x^\prime,t^\prime)
          \psi(x^\prime,t^\prime) - \eta(x,t) | ^2}{\sum\limits_{x\in L^3} |\psi(x,t) |^2},
\eeq
where the normalisation of the metric reflects the 
exponential fall off of the pseudoscalar correlator, 
was used to define the convergence criterion.
Using the variance of previous simulation data to estimate the
expected statistical error in the correlation functions, we found that requiring
the timesliced residue
on timeslice 24 ( the noisiest ) of $r_{t=24} \le O(10^{-11})$, 
gives an error due to convergence when comparing to 
an extremely tightly converged propagator that is 
orders of magnitude below the expected variance. This was subsequently
used as a convergence criterion and corresponds to a traditional residue
of between $10^{-8}$ and $10^{-12}$ depending on the quark mass, 
significantly tighter than the $10^{-7}$ obtainable using 32 bit precision.

Both local and fuzzed sources were used \cite{cmi_fuzz}.
Fuzzed smearing uses Michael-Teper fuzzed links to covariantly transport
the quark fields $n$ links along each of the axes to form a 
face-centred cubic smearing function.
The fuzzing radius $n = 4$ was selected to optimise the plateau for P-states using a
small number of configurations, which, in combination
with sink fuzzing, allowed a $2\times2$ smearing matrix to be constructed,
\beq
\label{op_matrix}
\langle {\cal O}_i {\cal O}_j \rangle = \left( 
\begin{array}{cc}
\langle {\cal O}_L(x) {\cal O}_L(0) \rangle & \langle {\cal O}_F(x) {\cal O}_L(0) \rangle\\
\langle {\cal O}_L(x) {\cal O}_F(0) \rangle & \langle {\cal O}_F(x) {\cal O}_F(0) \rangle
\end{array}
\right).
\eeq

\newcommand{\qq}{q\bar{q}}
\newcommand{\Qq}{Q\bar{q}}
\newcommand{\QQ}{Q\bar{Q}}
\newcommand{\BC}{Q\bar{c}}

Three additional propagators, from covariant derivative sources 
\beq
\Delta_i (x) = 
\sum\limits_y \left[ U_i(x) \delta(x+\hat{i},y) - U_i^\dagger(x-\hat{i}) \delta(x-\hat{i},y) \right] 
\delta(y,0),
\eeq
were generated at $\kappa=0.12600$, 
lying closest to charm,  allowing
the differential operators in \tab{tab:operators} to be constructed
for mesons composed of this quark and one at another value of $\kappa$.
The most important result of this is that we now have access 
to the angular momentum 2 P-state via the $2^{++}$ operator, in both the 
$T$ and $E$ representations.

\subsection{Mass Fitting of Correlators}

In what follows, all fits to correlators were made using correlated
fits with the Marquardt Levenberg algorithm, and
the bootstrap algorithm was used to estimate the errors,
using $1000$ bootstrap samples.
Single exponential, row fit and matrix fit models were used. 

The fit ranges were varied
with $t_{\rm max}$ holding typically three values
selected by inspection of the effective mass plots,
and $t_{\rm min}$ varied over the entire range in which the
Marquardt algorithm converged for the fit model.
This procedure was carried out for each operator ${\cal O}
 \in \{0^{-+},1^{--},0^{++},1^{++},1^{+-}\}$ and, where the data existed,
for the $2^{++}$ operator.

The information was used to select
an optimal fitting range by considering the stability of the central value and
error with varying $t_{\rm min}$, requiring $Q \ge {\cal O}(0.1)$ and 
$\chi^2/{\rm dof} \le {\cal O}(1.5)$, and considering the consistency in the 
fitted masses between the double and single exponential fit methods.

The simultaneous double exponential fit to the fuzzed-source local-sink, and the
local-local correlator was the selected method, and the
ranges in \tab{Tab60FitRanges} were deemed optimal. These were subsequently
used to obtain the best fit masses in Table \ref{TabHHSstates} and 
Table \ref{TabHHPstates}. The excited state masses given in these tables
were not found to be sufficiently independent of the fit range to produce
confident spectrum predictions.
The effective mass plots obtained for the various angular momentum
states of the $\kappa_1=\kappa_2=0.12600$
system are displayed in \fig{FigHHpsMeff} to \fig{FigQQFineDiffMeffs}, where the 
open circles are the points from the local-local correlation function, and the 
fancy diamonds from the fuzzed-local. 
They are considered representative of the plots
for other $\kappa$ combinations.

Since the P-states have a large splitting from the pseudoscalar,
the noise grows rapidly with time \cite{kronfeld_noise}, 
and it is essential to perform part 
of the fit at early timeslices to obtain a good signal, requiring a well chosen
fuzzing or smearing radius. 
\fig{FigQQFineDiffMeffs} clearly shows that we have resolved 
the complete fine structure of the $\chi$ states for the first time 
with a relativistic action, since the $68\%$ error bounds do not overlap.
We, of course, obtain a reduced error by taking 
the bootstrapped difference in the
fitted mass values.

\subsection{Kinetic Masses}
The correlation functions were computed at various momenta
allowing us to compute the kinetic mass by fitting the
dispersion relation 
\beq
E(p^2) = m_1 + \frac{p^2}{2m_2} + C p^4.
\eeq

The non-zero momentum behaviour of the $(2600,2600)$ pseudoscalar,
\fig{Fig2600M2Disp} is considered typical, and the fitted lattice kinetic masses are given in 
\tab{TabHHM2}.

\subsection{Scale from Quarkonium S-P splitting}
When quarkonia have been simulated (using either NRQCD or
the heavy Wilson approach) quenched calculations have obtained 
a large slope in the $\frac{1}{M}$  dependence of the spin-averaged 
$1S-1P$ splitting while the experimental splitting in 
$\Upsilon$ and $J/\psi$ is thought to differ by only a few MeV (with reasonable 
assumptions for the $\eta_b$ mass).
The slope calculated is a quenching effect \cite{ChristineTsukuba}.
It has been argued that the source of this 
slope is the incorrect running of the coupling in the 
quenched approximation \cite{khadra_SP_running,Aoki_Tsukuba}
and that the definition of  $a^{-1}$ appropriate for quarkonia
is that taken from the S-P splitting, in an attempt
to eliminate quenching effects inherent in setting the scale from 
a low momentum quantity.
This definition of the lattice spacing therefore depends on the
mass of the heavy quarks used to determine it. 
The mass dependence of the quarkonium $1P-1S$ splitting and
the corresponding inverse lattice spacing obtained in this  simulation is shown in
\fig{FigHAinvSP}.
The inverse lattice spacings at $\beta=6.0$ obtained for $J/\psi$ 
and $\Upsilon$ are $2.17(6)$ GeV and $2.4(1)$ GeV respectively.
The value obtained using NRQCD on the same lattice was $2.5(1)$ for
the $\Upsilon$ system \cite{cdavies_upsilon_long}.
The inverse lattice spacings obtained from $M_\rho$ ( with the 
tadpole improved action \cite{parlat97} )
and $\sqrt{K}$ are 1.96(4) GeV and 2.02(3) GeV respectively.

It has recently been suggested \cite{BaliBoyle} based on potential models
that the errors in the quenched quarkonium spectrum are dominated by
an overestimation of the $1S$ state relative to the physically larger states
due to underestimation of the Coulomb coefficient 
in the quenched potential. As a result,
even under a redefinition of the scale, the quenched quarkonium spectrum
is inconsistent with experiment. 
For these reasons we analyse the $\beta=6.0$ simulation in three ways,
using the $\rho$ mass, the string tension, and the quarkonium S-P splitting
to set the scale, the latter being included largely to allow comparison to
other calculations.
Finally we note that the sensitivity of both the hyperfine splitting
and the $S-P$ splitting to quenching makes them 
excellent candidates for observing unquenching effects in hadrons
in exploratory dynamical simulations.

\subsection{Charm and Beauty Systems}
The extrapolations in inverse heavy quark mass 
for the heavy-heavy systems were found to be amenable to a linear fit. 
The hyperfine splitting, S-P splitting and 
$2^{++}$ - $1^{++}$ - $0^{++}$ fine structure
were calculated and extrapolated to match
the kinetic spin-averaged S-state mass, $M_2$, to the physical mass,
setting the scale using each of the string tension, $M_\rho$,
or the S-P splitting. 
Here we artificially assume that the (experimentally unobserved) 
$\eta_b$ mass lies $40$ MeV below $\Upsilon$ 
assuming linear fall in the hyperfine splitting from $J/\psi$ with $\frac{1}{M}$.
This will introduces at most an $O( 1 / 1000  )$ error in the bare quark mass
after fixing the mass of the spin-averaged S-state to the $\Upsilon$ system.

\subsubsection{Hyperfine Splitting}
A linear potential model argument \cite{Lucha}
gives the result that the quantity $(M_v^2 - M_{ps}^2)$ is constant,
and we compare the simulation data using $M_2$  to define
the $(M_v+M_{ps})$ factor in \fig{FigM1Hyperfines}. Here we have included
the light and heavy light data points obtained by UKQCD \cite{pboylelat97} on the 
same lattices, and we use $M_1$ for the light data points.

Here $M_\rho$ has been used to set the scale so that the 
data are normalised to experiment using $M_\rho^2 - M_\pi^2$,
and the overall shape is significant.
The experimental points are approximately constant
for the light and heavy-light systems, while
lying slightly higher for $J/\psi$. This probably results from 
the quarkonium S-states being significantly determined by the Coulomb part of the potential.
The simulation data are very flat above the $D$ meson 
and the quarkonium points do not demonstrate the  ``emergent'' behaviour
shown by experiment, which we attribute to the Coulomb pole increasing the
wavefunction at the origin. 
Since the data is normalised to experiment at the light quark end of the plot,
it appears that in the quenched approximation the hyperfine splittings
fall disproportionately quickly with quark mass when 
compared with the real world in
the $\pi$, $K$, $D$ sector. 

\fig{FigHH_hyperfine60} presents the extrapolation of the quarkonium hyperfine 
splitting to charm and beauty using the
string tension to scale both axes in the extrapolation. The mass dependence of the splitting is
a straight line through the origin, in agreement with the non-relativistic 
expansion of QCD, and
in contrast to the extrapolations obtained with the tree-level clover and unimproved 
Wilson actions,
\cite{sarathesis,UKQCDhyperfine}, as seen in \fig{saraQQhyp}. 
The size of the hyperfine splitting is in clear disagreement 
with that in  the $J/\psi$ system. 

\subsubsection{Fine Structure}

We compute the P-states using the $\gamma$-based operators where possible; they
were found to be significantly statistically cleaner, as can be seen by comparing
figures \ref{FigQQFineMeffs} and \ref{FigQQFineDiffMeffs}. Furthermore data existed for the degenerate
quarkonium with the $\gamma$ operators for each of the $\kappa$ values.
The $2^{++}$ state, however, required extra derivative source propagators for one of the quarks in
the meson; these were generated only for
$\kappa=0.12600$, so that the fine splittings involving the $2^{++}$ state were
extrapolated linearly using non-degenerate mesons, to the point where the non-degenerate
pseudoscalar inverse kinetic mass was equal to the $\eta_c$ mass. We generated correlation functions
for operators in both the $T_2$ and $E$ representations and found them to be degenerate 
within our statistical error. We take the fitted mass from the $T_2$ representation
which we found to be statistically slightly cleaner.

This analysis is not ideal
but errors introduced may be estimated using the non-degenerate Hamiltonian in \cite{Lucha2}.
This is easily re-written in terms of the the combinations $(m_1 + m_2)$ and $(m_1 - m_2)$
giving that the corrections to the fine structure Hamiltonian for non-degenerate quarks are 
proportional to $\frac{m_1 - m_2}{m_1 + m_2}$. In extrapolating to charm from the
nearby $\kappa=0.12600$, the correction to the fine splitting is of order $10\%$,
which is lower than the level of statistical error. 
Extrapolations to the beauty systems via this method however are problematic.

The extrapolations are performed using degenerate quarks and the 
$\gamma$-based operators for splittings not involving the $2^{++}$ state.
The $m_Q$ dependence of the fine structure
is shown in Figures \ref{FigHH_Xc0_Xc1fine60}, 
\ref{FigHH_Xc1_Hcfine60}, \ref{FigHH_Xc2_Hcfine60} 
and \ref{FigHH_Xc2_Xc0fine60}. As can be seen, we resolve
completely the $\chi$ triplet, however, the calculated splitting between
the $h_c$ and $\chi_{c1}$ is consistent with zero.
We obtain estimates for the ratio 
$\frac{M(^3P_2)-M(^3P_1)}{M(^3P_1)-M(^3P_0)}$ of 
1.6(6) for charmonium and the systematics 
in extrapolating to bottomonium leave the extrapolation not worth
considering here. The experimental value is $0.48$ for charmonium and $0.66$
for bottomonium. The same splitting when calculated in 
NRQCD has been shown to suffer from significant 
discretisation effects \cite{ChristineTsukuba}, so that 
without calculation of the same quantity at other lattice spacings we cannot
distinguish between discretisation and quenching effects.

\subsubsection{Spectrum Predictions}

The results obtained using each of the string tension, the $\rho$ mass,
and the quarkonium S-P splitting,
extrapolated to both the $J/\psi$ and $\Upsilon$ systems can be found in \tab{Tab60Quarkonium}.
As an illustration of the resulting data we present diagrams for the spectrum obtained using
the string tension for both the $J/\psi$, \fig{FigHH_Spectrum}, and the $\Upsilon$,
\fig{FigHH_UpsSpectrum}, systems.
The spectrum obtained for $J/\psi$ is broadly comparable to experiment,
with the spin splittings underestimated, while we expect
$\Upsilon$ the results show
severe quenching effects in both the hyperfine and $S-P$
splittings. 

Our values for the hyperfine splitting in charmonium 
however, agree with the FNAL quenched data which has been
included in reviews by Shigemitsu \cite{JunkoLat96} and Davies \cite{ChristineTsukuba}.
For both systems the discrepancies in the fine structure
may well include strong discretisation effects. It is common for the lattice spacing 
to be set independently in each of the $J/\psi$ and $\Upsilon$ systems via
the $S-P$  splitting at that quark mass. This would certainly improve the 
spectrum diagram we obtain for $\Upsilon$, however it is felt that, since
this defines a different quenched theory for each system, this obscures the
fact that the one quenched theory cannot simultaneously reproduce the spectra of
these systems.
Previous calculations of charmonium P-states with a relativistic lattice
action have been made \cite{rajan,pboylelat96,pboylelat97,Mackenzielat97}, however, this is
the first relativistic lattice calculation that competely resolves the $\chi$ triplet.

\section{Scaling Behaviour of Hyperfine Splitting}

\label{CHAPTERSCALING}
We also performed a less extensive 
calculation using the tadpole-improved clover action at $\beta=6.2$,
where we consider only the S-states.
The simulation used 150 sample gauge configurations from
a quenched distribution at $\beta=6.2$ on a $24^3\times48$ lattice.
Five heavy quark masses corresponding to,
\beqau
\kappa_{\rm heavy} \in \{0.12000, 0.12300, 0.12600, 0.12900, 0.13200\},
\eeqau
were used,
all having the tadpole improved clover action with $C_{SW} = 1.44239$, corresponding to 
$u_0 =  0.88506$ taken from the plaquette.
The data were fitted using a similar procedure to that described previously.

With only two calculations, an extrapolation to the continuum using a linear model
is ill-advised, and so we plot
the lattice spacing dependence without performing a fit. 
The scaling behaviour of the $J/\psi$ hyperfine splitting
is shown in \fig{FigUpsHyperfineContinuum}, and can be
seen to be near scaling with both the string tension and $M_\rho$ scales,
in clear disagreement with experiment.
The charmonium hyperfine results 
are consistent with those of El-Khadra et al \cite{kronfeld_mflgt}.
Based on this it is reasonable to assume that our results on the $\Upsilon$ hyperfine
splitting underestimate the (as yet undetermined) experimental value.

Future calculations should produce a result for 
the hyperfine splitting which scales using the quarkonium S-P
splitting to set the lattice spacing. This will probably not agree with those using
the low momentum observables since NRQCD calculations
produce a result \cite{ChristineTsukuba}
for the S-P splitting in units of $M_\rho$ which scales and is 30\% lower than experiment for $\Upsilon$;
this induces an inconsistency between the quarkonium hyperfine splitting
scaled with each of these quantities.
The mass dependence of the quenched S-P splitting will, of course, mean that
the level of discrepancy between the lattice spacing definitions will differ in the $J/\psi$ and
$\Upsilon$ systems. 
For comparison we can estimate the value NRQCD calculations of the hyperfine splitting
would yield using $M_\rho$ to set the scale 
by assuming $\Delta_{\rm hyperfine} \propto \frac{1}{m_Q}$;
feeding the change in scale into both the selected quark mass and the splitting
then yields a 40-50\% reduction in the hyperfine splitting.
The NRQCD quenched continuum result appears to be around
$35$ MeV \cite{ChristineTsukuba} using the S-P splitting to set the scale,
with a 30\% uncertainty due to $u_0$ ambiguities and
higher order relativistic corrections. This gives an estimate for the central value
of around 17-21 MeV if $M_\rho$ were used to set the scale.

\section{Charm Quark Mass}
We first fix the bare lattice 
mass, $\kappa_{charm}$, by requiring the kinetic mass of the
pseudoscalar matches that of the $\eta_c$ using the string tension
to set the scale.
We then obtain values for the pole mass of the charm quark
using two methods \cite{kronfeld_lat97} involving either $m_1$ or $m_2$, which differ due
to lattice spacing effects. 

The first method evaluates the binding energy at that bare quark
mass:
\beq
aB_1 = aM_1 (Q\bar{Q}) - 2 aM_{1Q}^{PT},
\eeq
where the perturbative pole mass contains its
one loop correction obtained from \cite{kronfeld_self_energy}
\beq
aM_{1Q}^{PT} = \sum\limits_{l=0}^{1} g^{2l} aM_1^{[l]}
\eeq
The binding energy, defined via $M_1$, is then subtracted from the
physical meson mass to give the charm pole mass.
\beq
M_{\rm ch}^{\rm pole} = M(1S) -   B_1
\eeq

The other method employed relates the one loop kinetic mass to the bare
quark mass using the all orders in $a m_Q$ renormalisation constant 
\cite{kronfeld_self_energy}.
\beq
M_2 =  Z_{M_2} \times m_2(M_1^{PT}),
\eeq
where $M_1^{PT}$ contains its one loop correction and
\beq
m_2(aM_1) = e^{aM_1} \frac{\sinh{aM_1}}{1+\sinh{aM_1}}
\eeq

The largest source of error in the calculation is the coupling
$g^2$. Ideally we use the Brodsky-Lepage-Mackenzie scale 
$q^*$ for the self-energy.
We have not however evaluated this scale and instead proceed in a similar
manner to \cite{gupta_masses,allton_masses} by varying the scale at which
we evaluate $\alpha_V(q)$ over the range $\frac{1}{2a} \le q \le \frac{2}{a} $
to estimate the likely higher order error.

The scaling behaviour obtained for the two methods for obtaining the
charm pole mass is given in \fig{FigCharmMass}. The error is dominated 
by the uncertainty in the coupling, however it is plausible that a
consistent continuum limit is reached by the two methods, and 
suggest the likely range for the pole mass is
$1.25$ GeV $\le m_{ch}^{\rm pole} \le 1.55$ GeV.
We find the pole mass is, naturally, rather insensitive to the upper
limit of $q$ and strongly sensitive to the lower limit on $q$ due to
the nature of the running coupling.

We note that the perturbative errors are large here,
in particular Gray et al \cite{broadhurst} find a large $O(\alpha_s^2)$ correction
in the relation between the pole and $\overline{MS}$ masses. Due to this uncertainty
we do not convert our results to $\overline{MS}$ and instead quote only the 
pole mass. This is not to say, of course, that we  believe
that the pole mass is better defined, but rather we are reluctant to apply the
one loop formula when a large two loop correction may exist.

\section{Conclusions}

The quenched hyperfine splittings in quarkonium systems scale
within statistical error between the two calculations
with either the string tension or $M_\rho$ used to set the scale, 
and are in clear disagreement with experiment for $J/\psi$ and with
phenomenological model predictions for $\Upsilon$. Our results for the
hyperfine splitting in charmonium are in agreement with \cite{kronfeld_mflgt},
and those for $\Upsilon$ are in plausible agreement with NRQCD results
when a correction for the different inverse lattice spacings used 
is made. Systematic errors are estimated to be at the 5\% level.

We make the first lattice calculation of the complete  $1P$
triplet fine structure
using a relativistic quark action, and find that the 
inconsistencies of the spectrum with experiment
in the calculation at $\beta=6.0$, namely the low hyperfine 
splitting and the mass dependence of the $S-P$ splitting are
consistently explicable by an underestimation of the Coulomb 
coefficient at hadronic length scales in the quenched approximation.
We give an estimate for the pole mass of the charm quark, with an error 
which is dominated by perturbative uncertainties.

\section{acknowledgements}
We acknowledge the support of EPSRC grant  GR/K41663 and PPARC
grant GR/K55745. The calculations were performed using UKQCD time
on the Cray T3D in the EPCC at the University of Edinburgh.
Peter Boyle was funded by the Carnegie Trust for the Universities of Scotland
while this work was carried out at the University of Edinburgh,
and is grateful for PPARC grant PP/CBA/62 and both the University of Glasgow,
and University of California, Santa Barbara, where this work was written up.
I acknowledge many useful conversations with Christine Davies during the course
of this work.


%
%

\pagebreak

\fltfig{Estimated error in mean kinetic energy estimated at tree level as
a fraction of the total kinetic energy, $\Delta_{\rm kin}$ at $\beta=6.0$ and
$\beta=6.2$ due to discretisation effects. We used an approximation
to  potential model values $\langle v^2 \rangle$ in this estimate.
Estimated systematic errors are of the order 5\%.}
{\pspicture{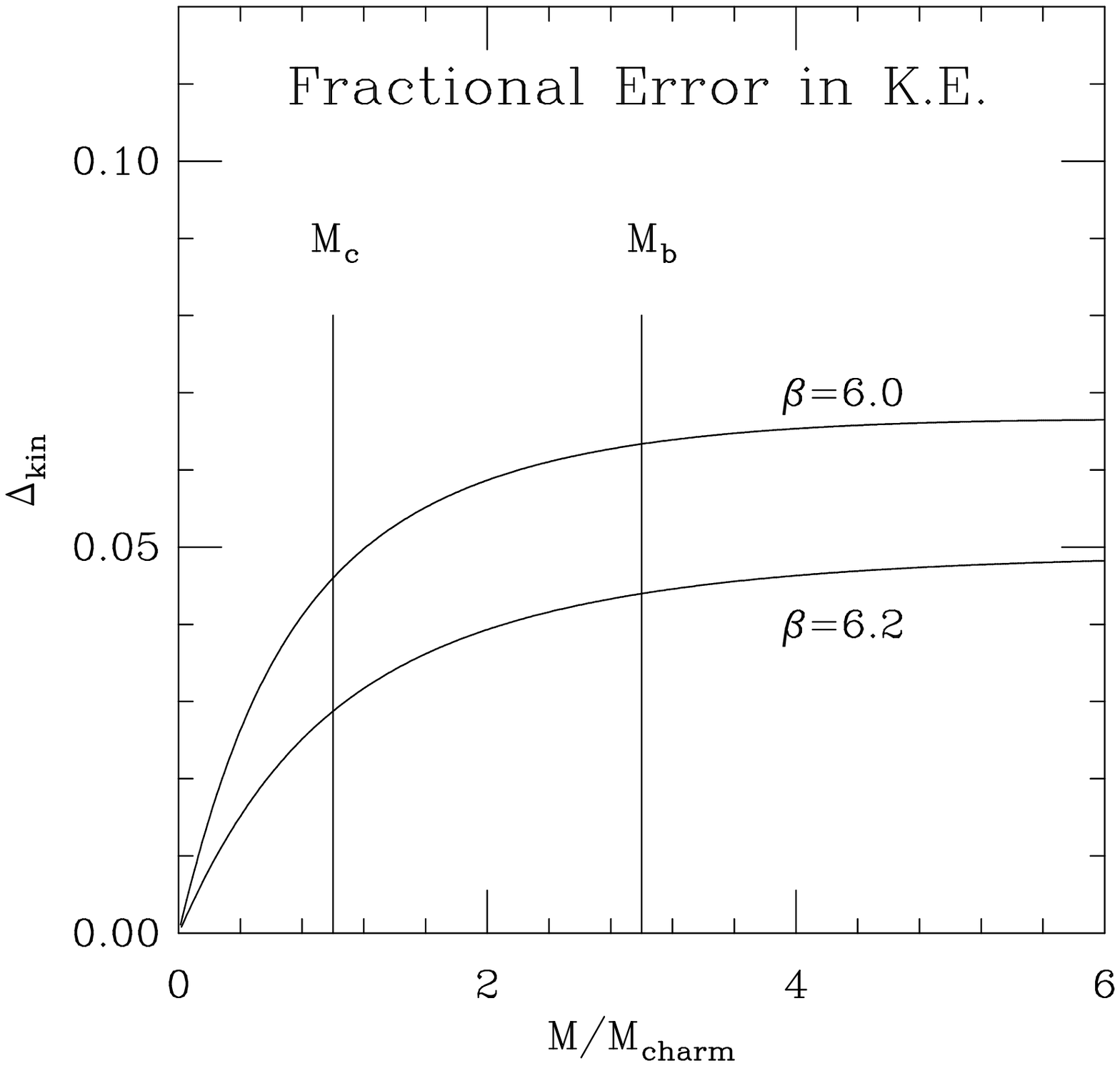}}
{FigKineticError}

\fltfig{Heavy-heavy pseudoscalar and vector effective masses for a 
meson containing degenerate quarks with $\kappa=0.12600$ at $\beta=6.0$. These plots
are considered representative of the rest of the data set. The hyperfine
splitting is clearly seen. The circles are points from local-local correlation
functions, and the fancy squares come from the fuzzed source, local sink correlation
functions.}{
\epsfaxhax{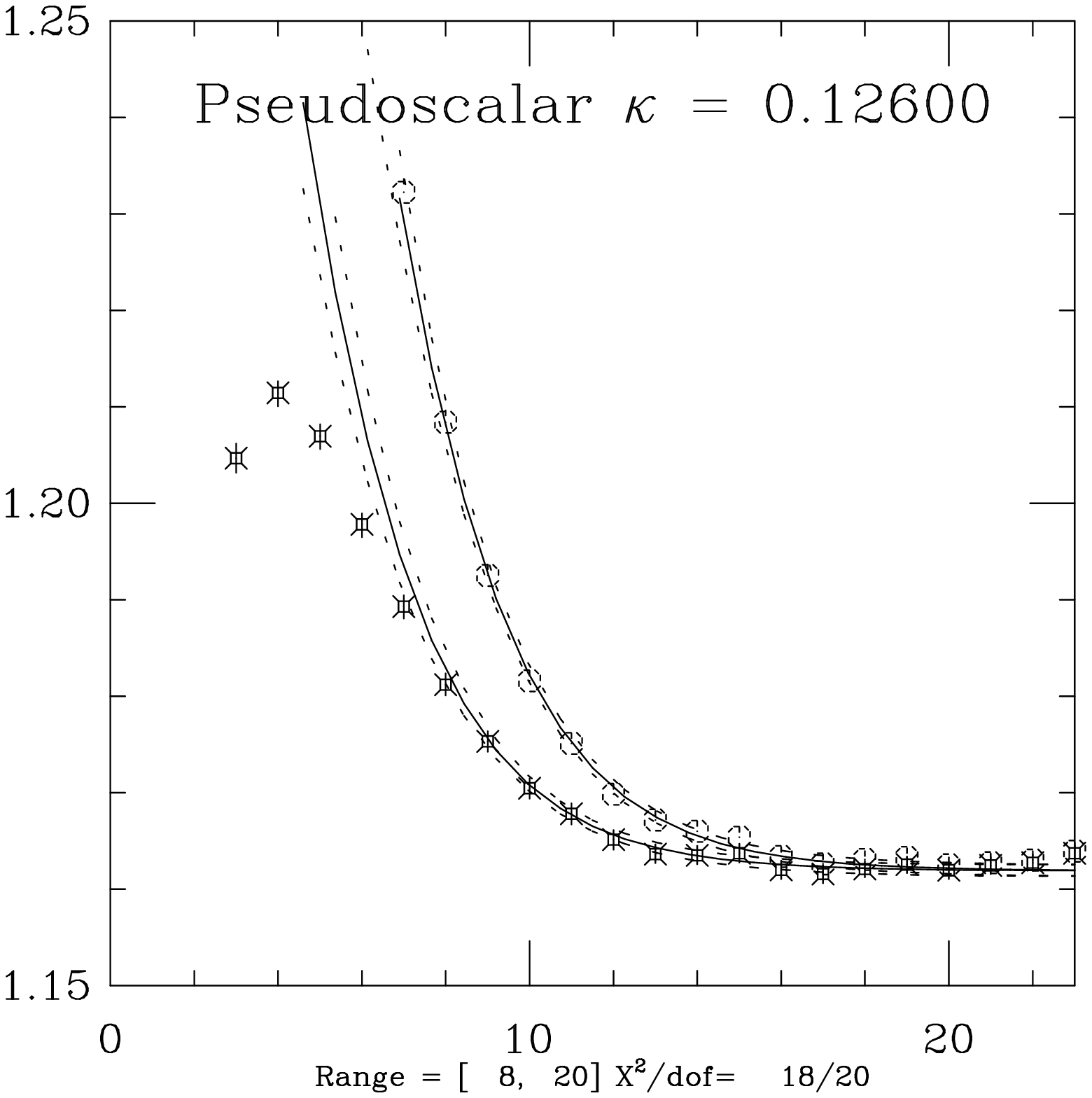}
{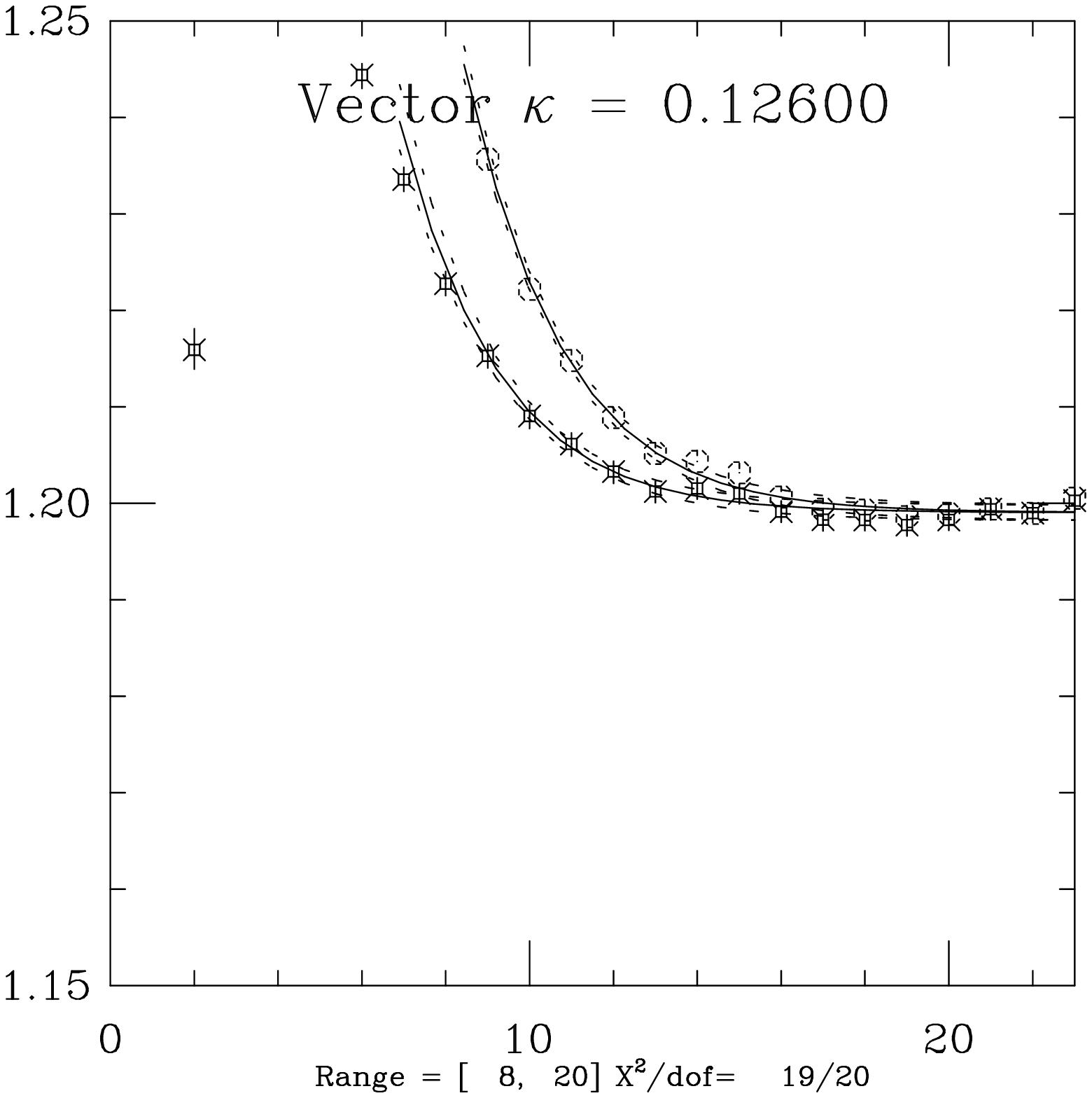}
}{FigHHpsMeff}

\fltfig{Quarkonium $\chi_0$,$\chi_1$, and $h$ states using $\gamma$ operators for a 
meson containing degenerate quarks with $\kappa=0.12600$ at $\beta=6.0$. The $0^+ - 1^+$
splitting can be clearly seen. The $1^{++}$ and $1^{+-}$ states are not
resolved. The circles are points from local-local correlation
functions, and the fancy squares come from the fuzzed source, local sink correlation
functions. }{
\epsfaxhaxhax{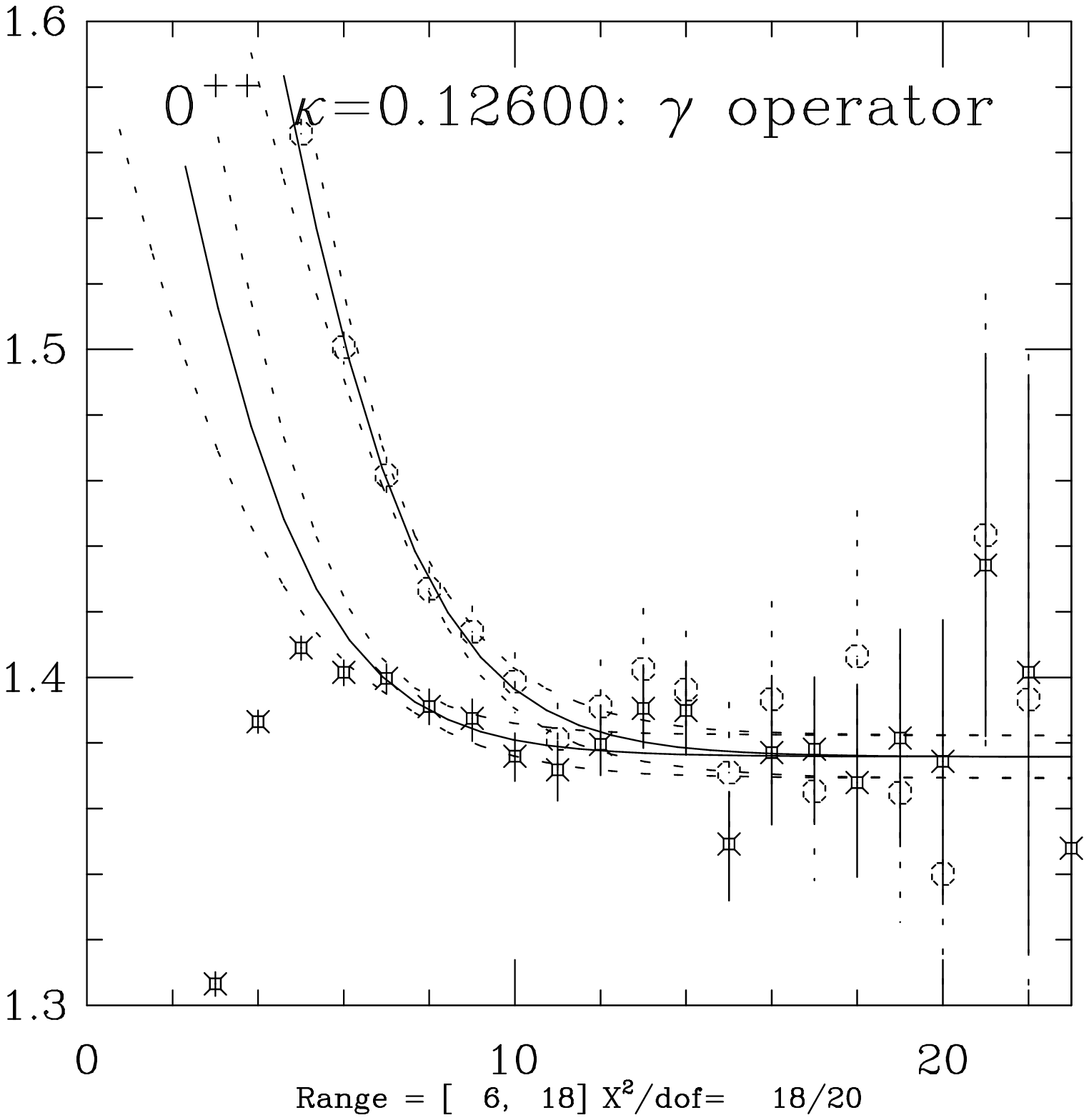}
{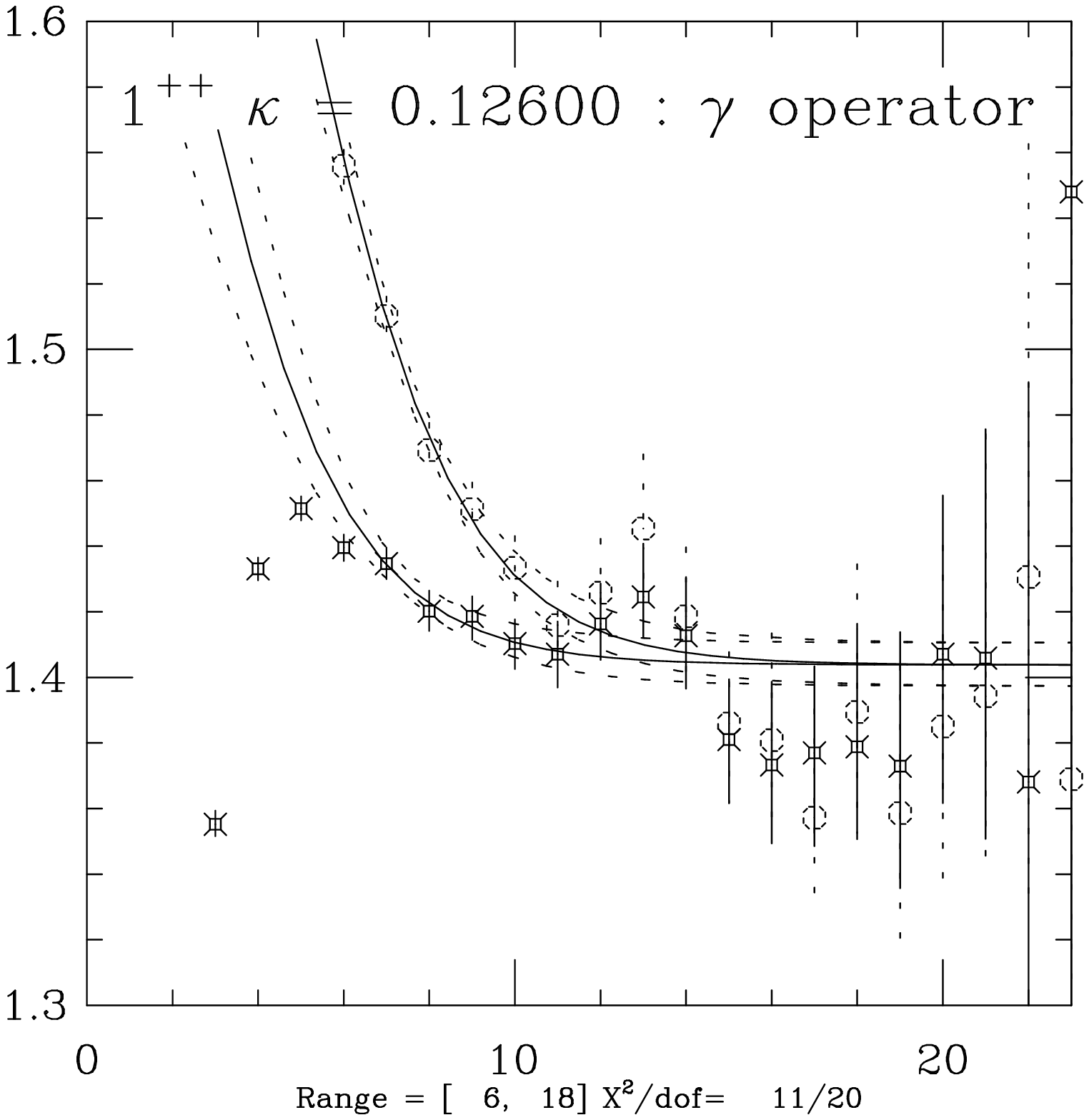}
{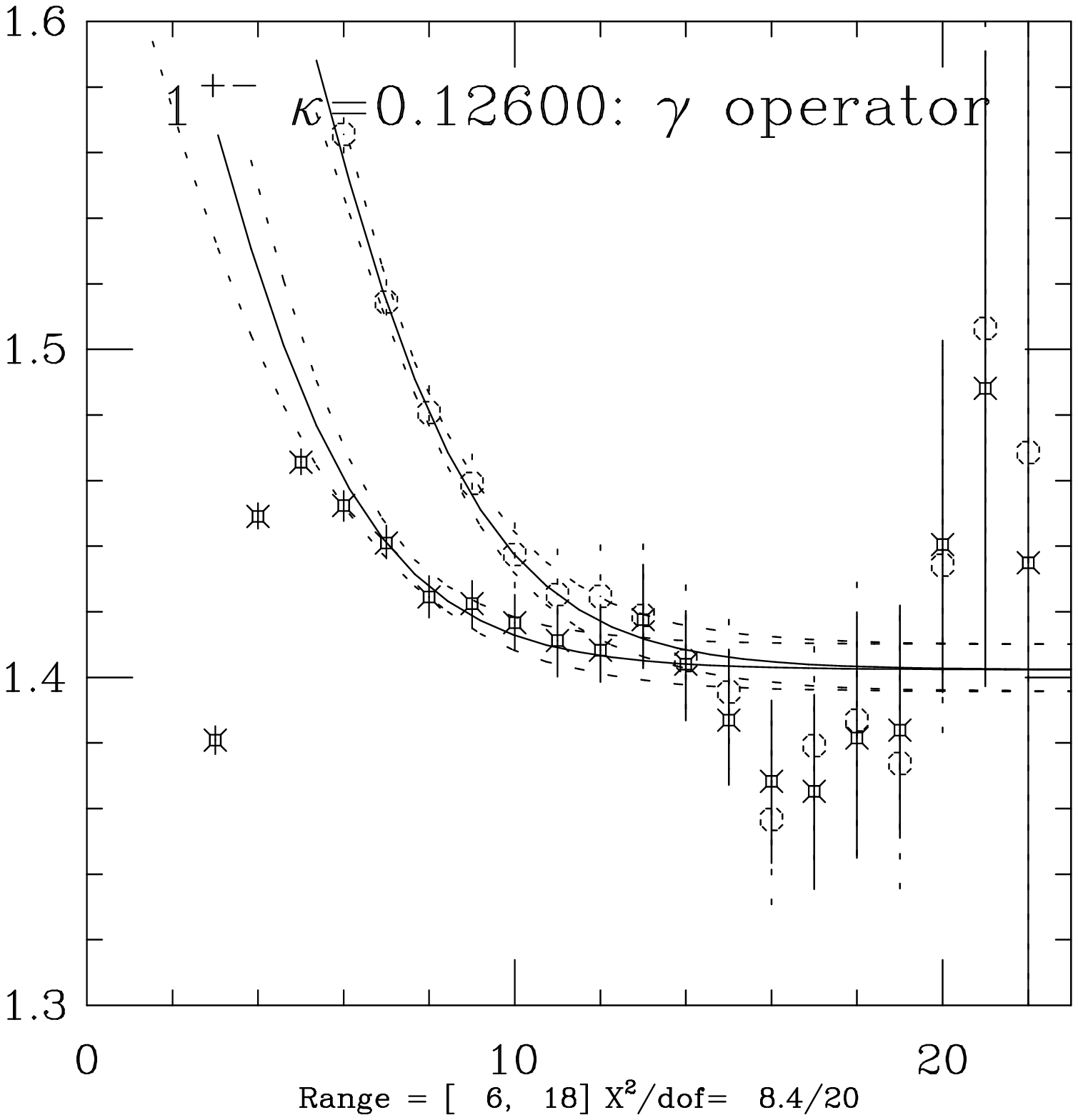}
}{FigQQFineMeffs}

\fltfig{Quarkonium $\chi_0$,$\chi_1$, and $\chi_2$ states using differential operators
for meson containing degenerate quarks with $\kappa=0.12600$ at $\beta=6.0$. 
The complete fine structure of the $\chi$ triplet is clearly resolved, even
before correlated differences are taken.}{
\epsfaxhaxhax{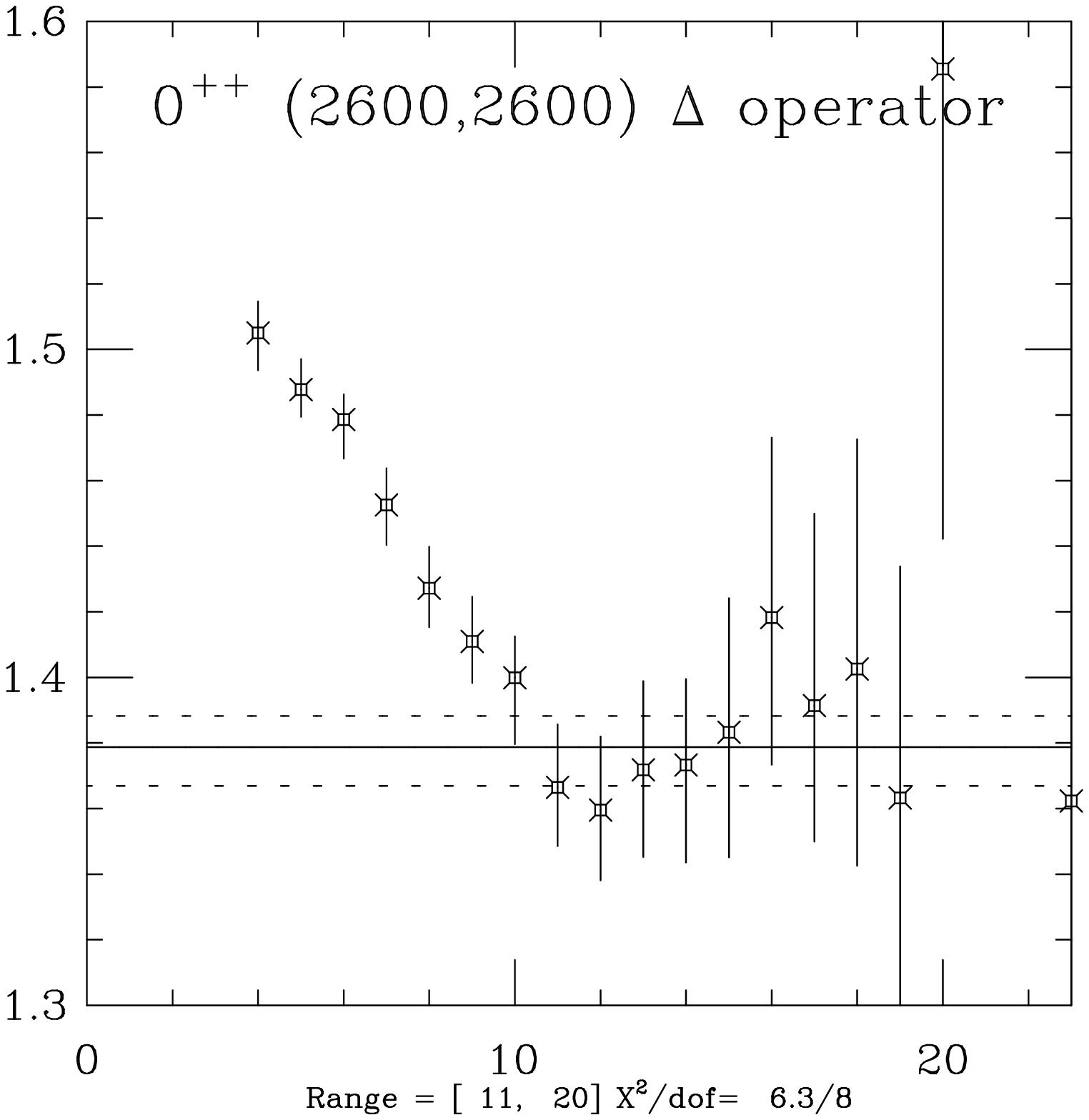}
{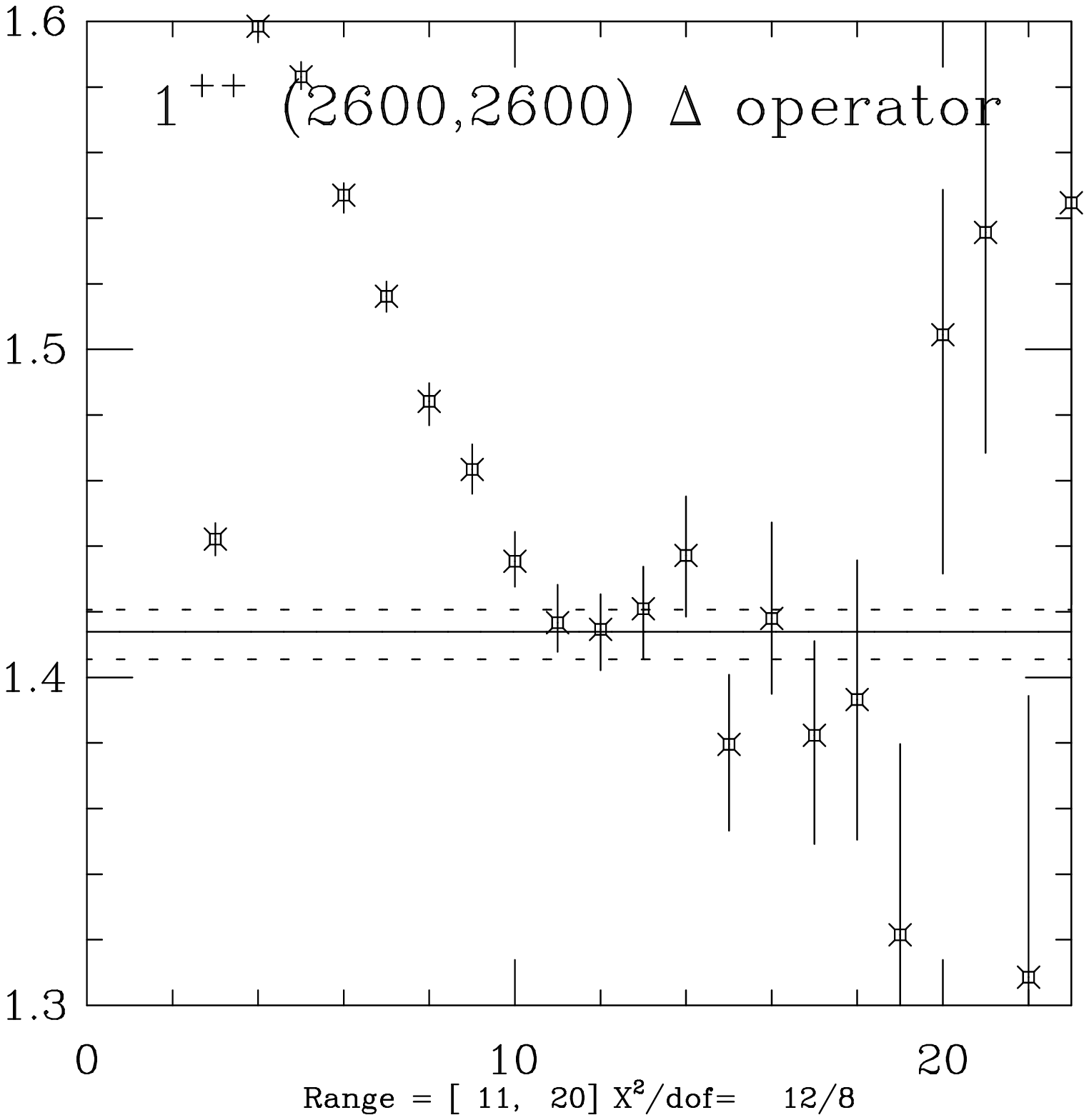}
{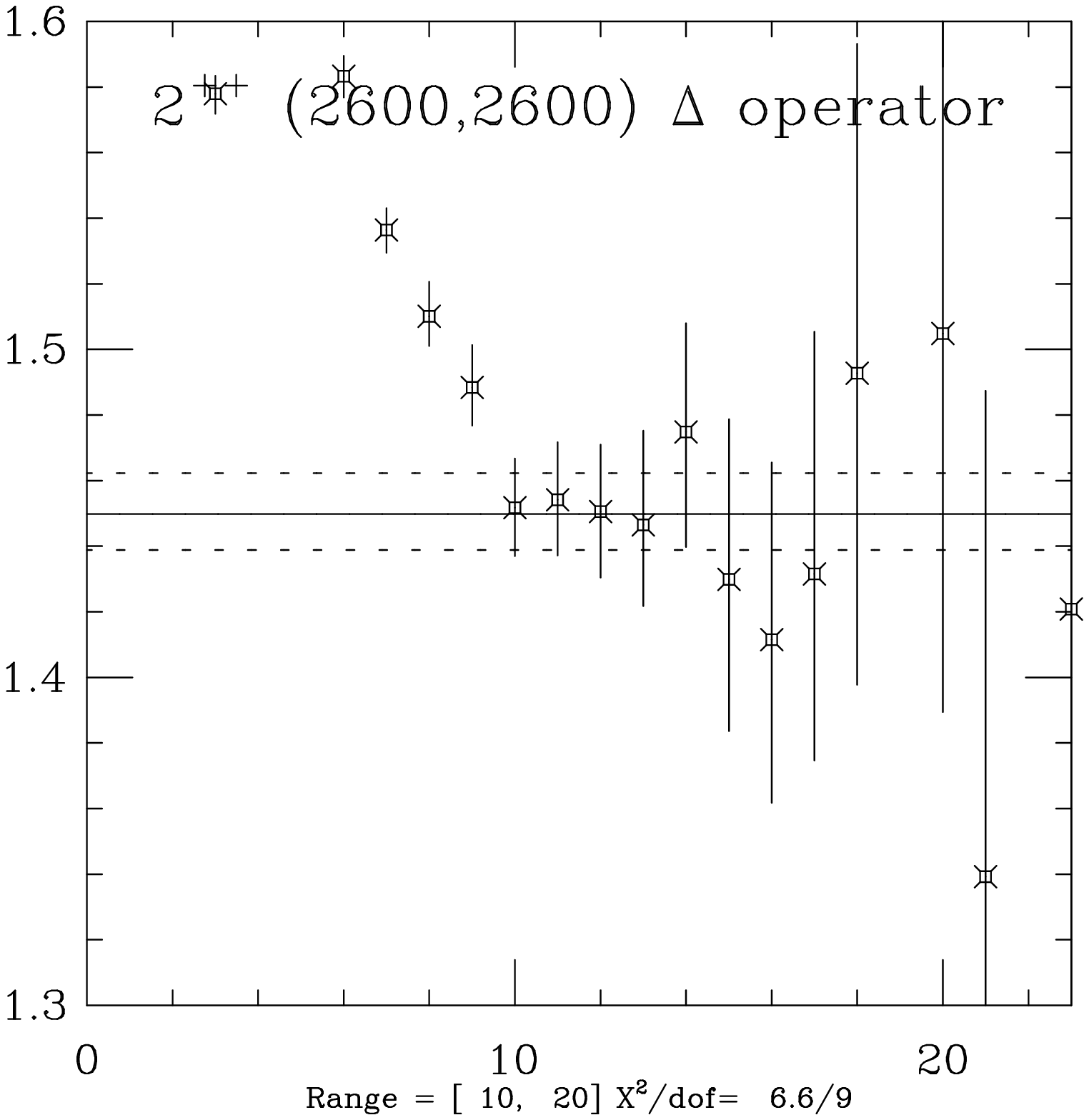}
}{FigQQFineDiffMeffs}

\fltfig{ Fit to the dispersion relation for the pseudoscalar meson containing 
degenerate quarks with $\kappa=0.12600$ at $\beta=6.0$. 
The fit is quadratic, and a significant
difference between the fitted $p^2$ coefficient and the naive derivative obtained
from the first two points is found, making the quadratic fit necessary.}{
\pspicture
{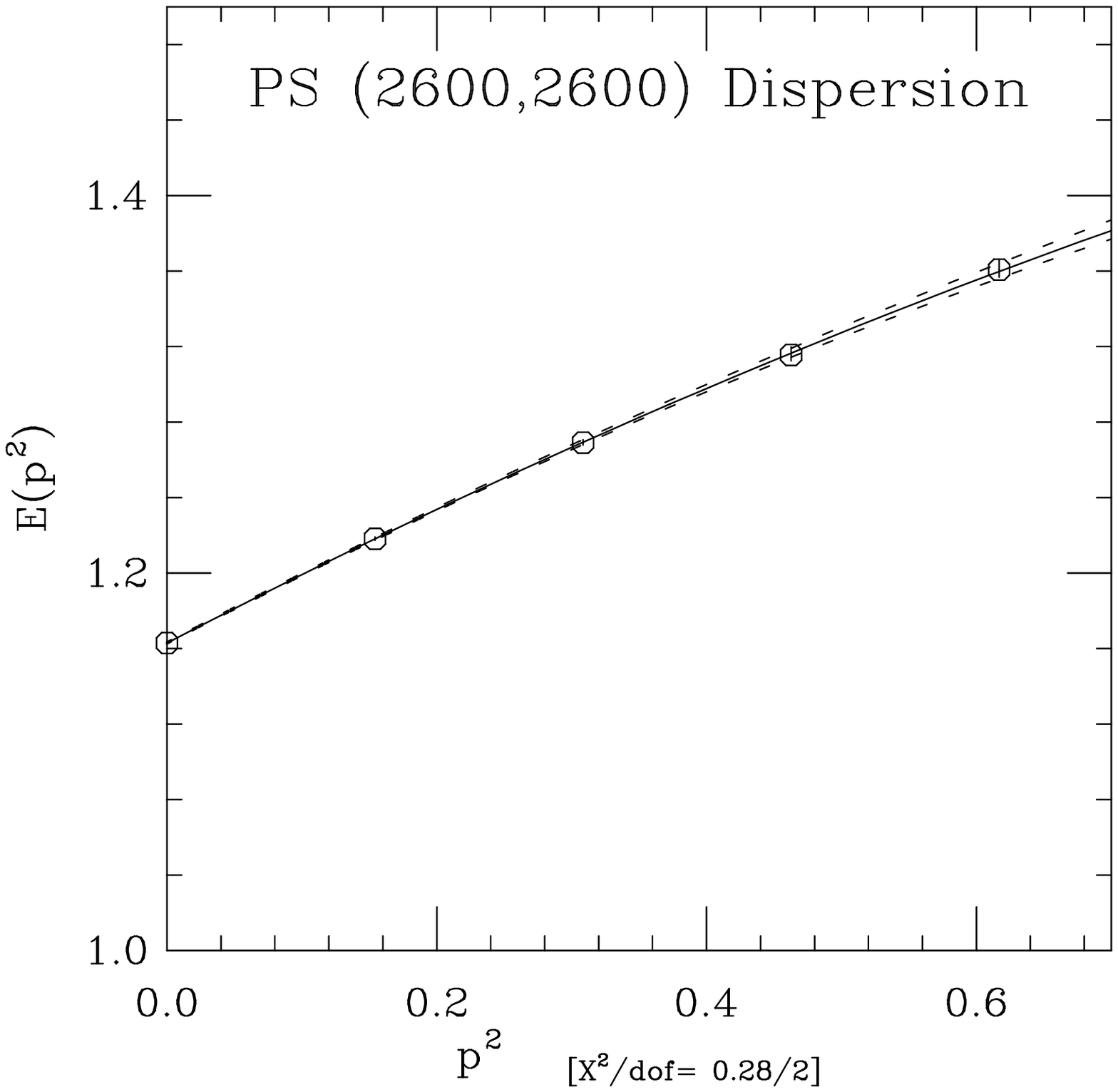}
}{Fig2600M2Disp}

\fltfig{Plots show the $S-P$ scaled to the string tension at $\beta=6.0$, and
the inverse lattice spacing obtained from quarkonium $S-P$ splitting as a function
of the inverse kinetic pseudoscalar meson mass. No significant deviations
from linearity are found within statistical error.}{
\epsfaxhax{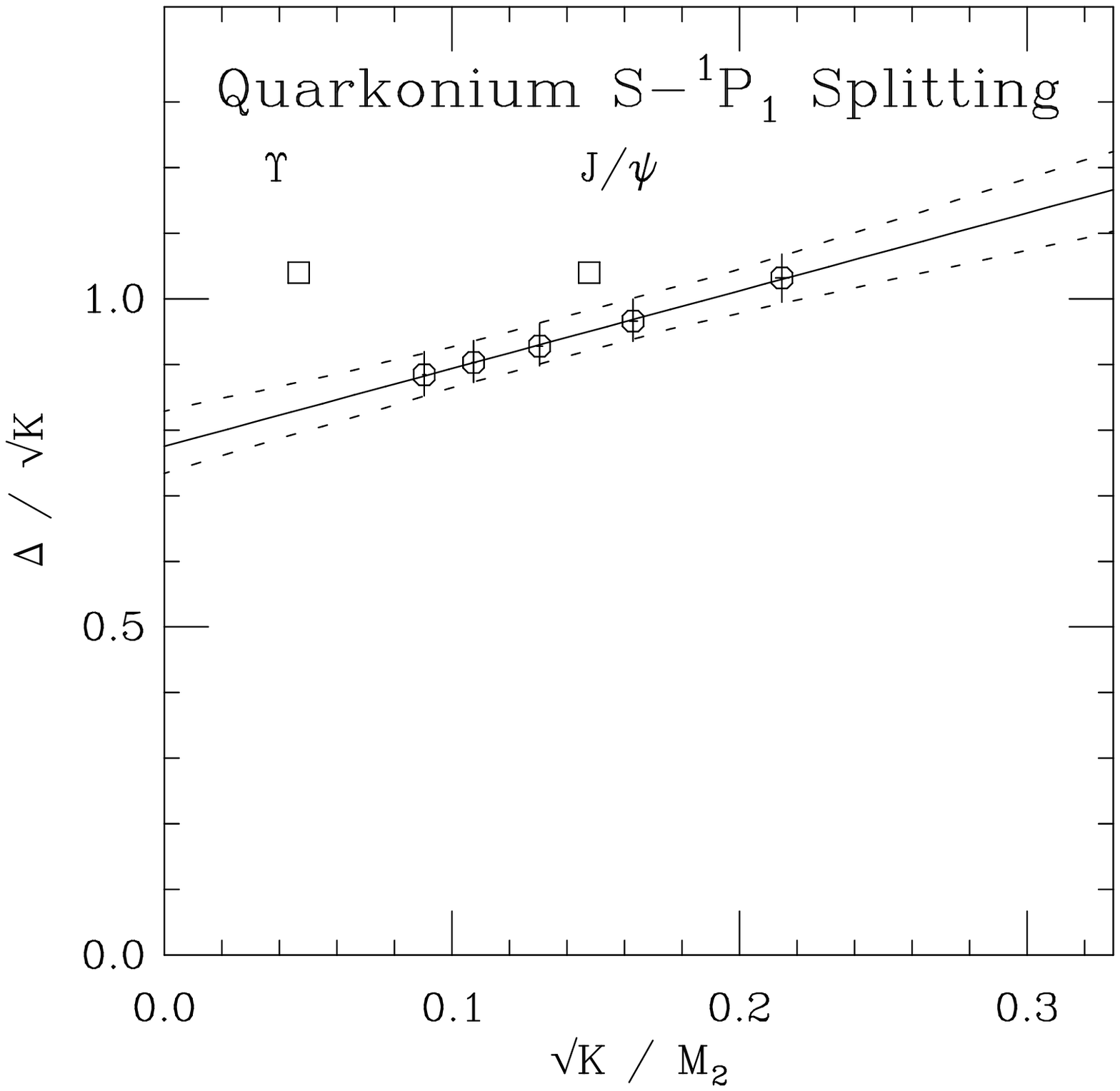}
{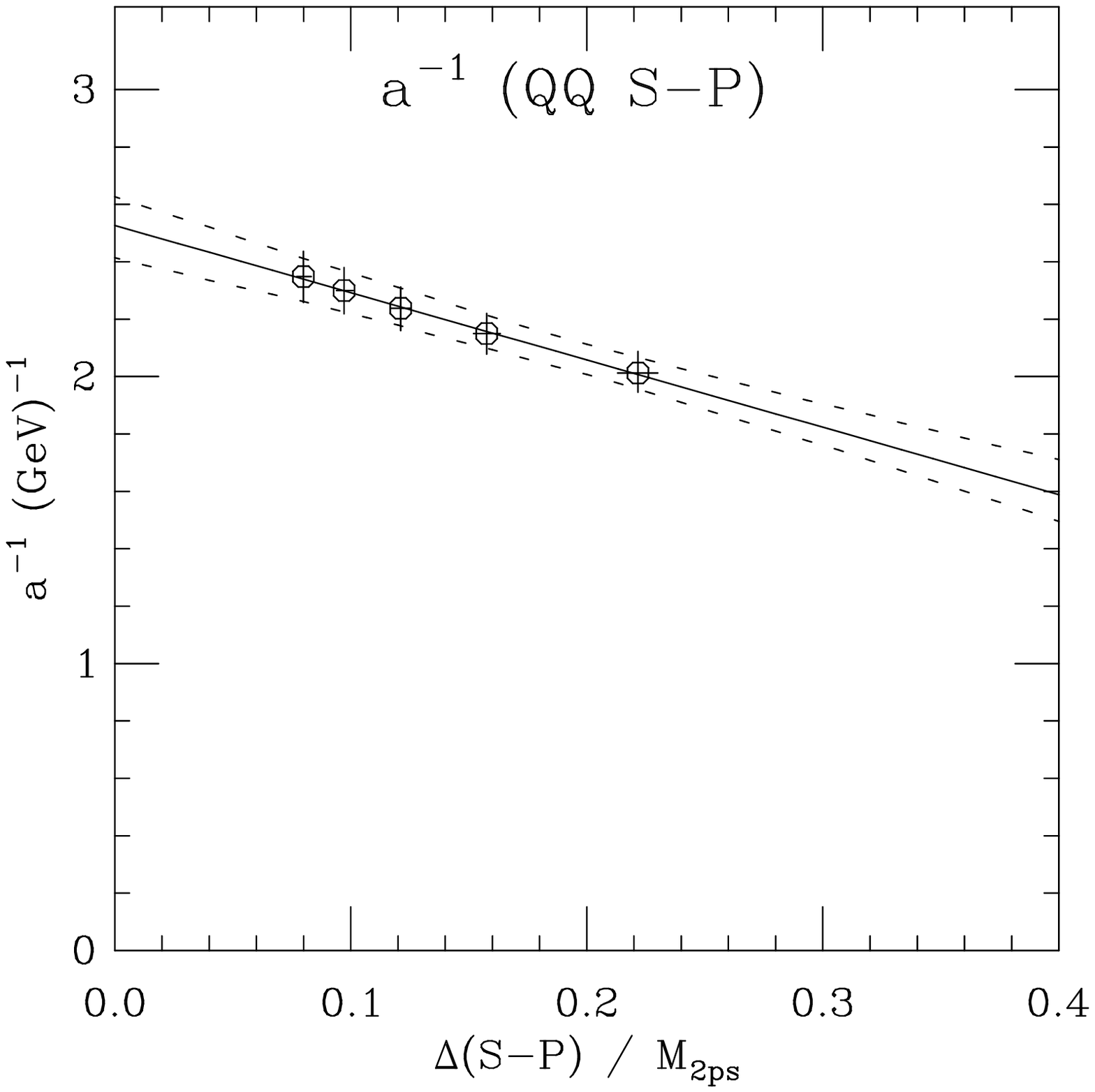}
}{FigHAinvSP}

\fltfig{Hyperfine splitting across entire mass range at $\beta=6.0$. 
The first three data points
are degenerate light quarks (open circles), there are fifteen heavy-light data points 
(fancy diamonds), and five degenerate heavy-heavy data points (fancy squares).
The heavy-heavy points show no sign of a rise relative to heavy-light points,
in contrast to the experimental behaviour.
This shows a clear defect of the quenched approximation. 
Here we have factorised $m_{\rm V}^2 - m_{\rm PS}^2$ and used a
kinetic mass factor $m_2(PS) + m_2(V)$ and a pole mass factor $m_1(V) - m_1(PS)$ 
for all but the light quarks, for which we use solely $m_1$ terms.
}
{
\pspicture
{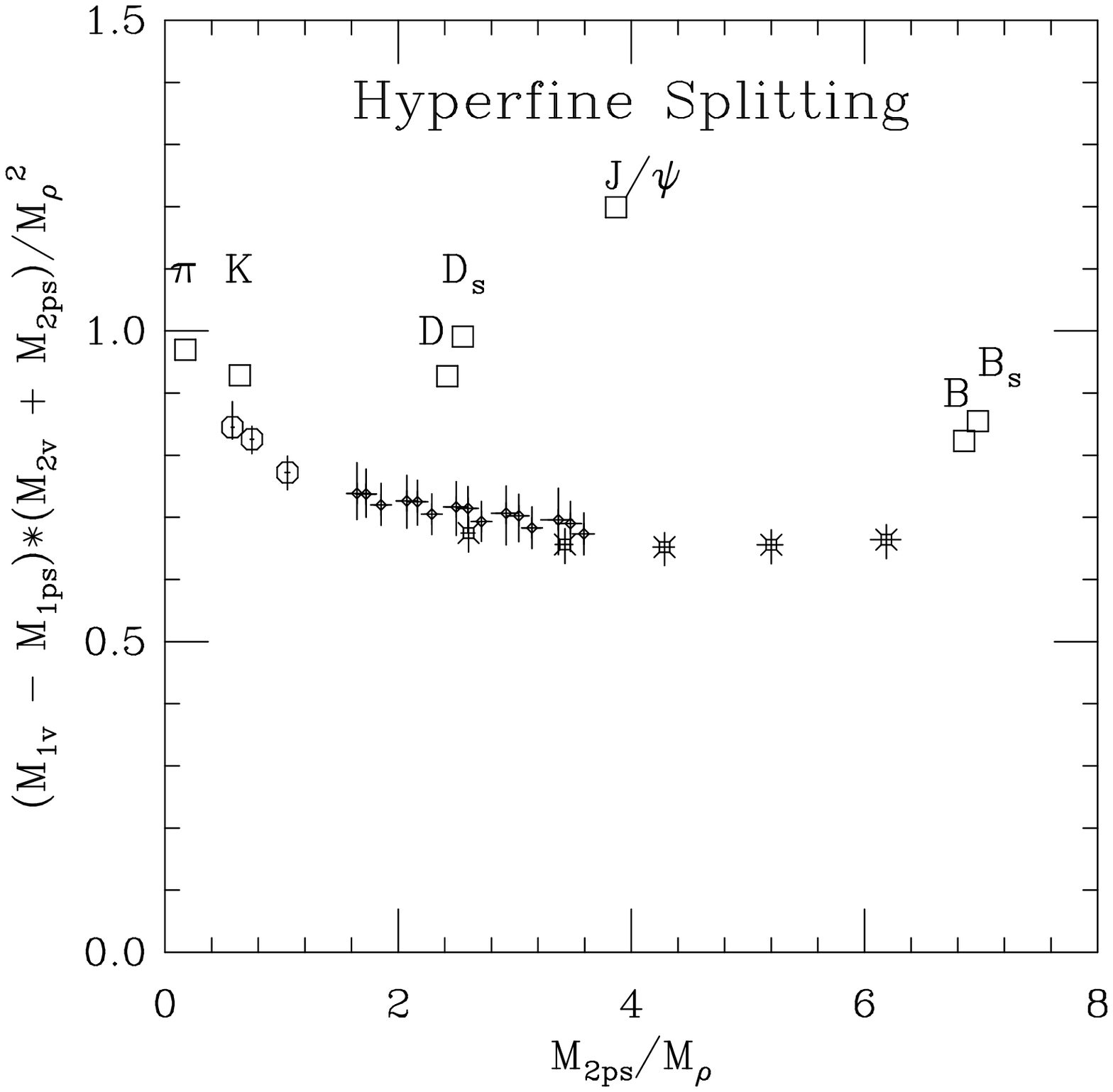}
}
{FigM1Hyperfines}

\fltfig{Extrapolation of quarkonium hyperfine splitting in the
inverse pseudoscalar meson mass at $\beta=6.0$. The axes in this plot are scaled to
the string tension. The ratio of the hyperfine splitting to the string tension
is clearly underestimated by the data compared with experiment. The hyperfine
splitting extrapolates to zero in the static limit. The axes are scaled using the string 
tension.
}{
\pspicture{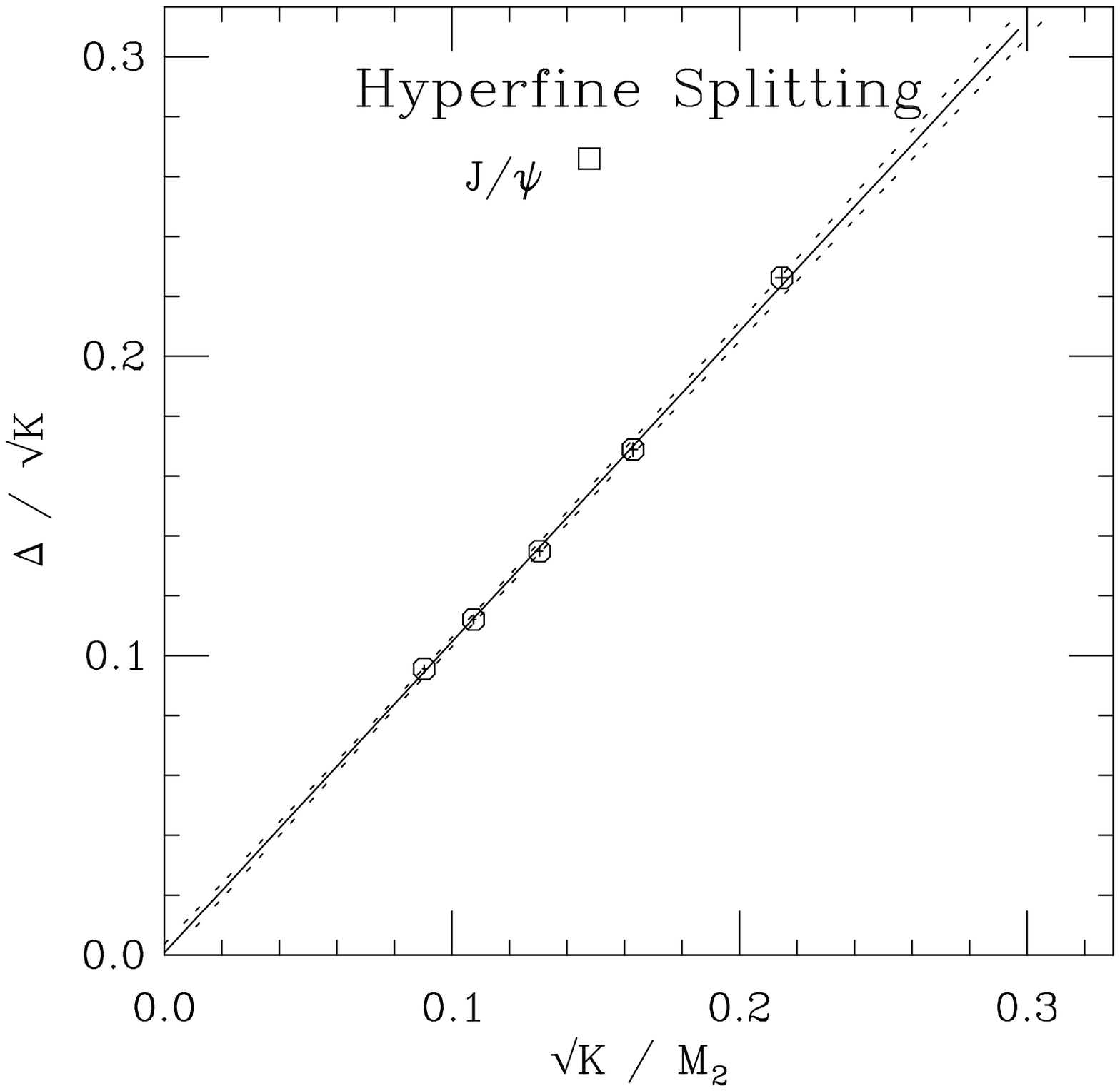}
}{FigHH_hyperfine60}

\fltfig{Previous UKQCD heavy quark extrapolation of quarkonium hyperfine splitting
using various actions. The hyperfine splitting clearly disappears before the static limit, in contrast
to the tadpole improved data set.}{
\pspicture{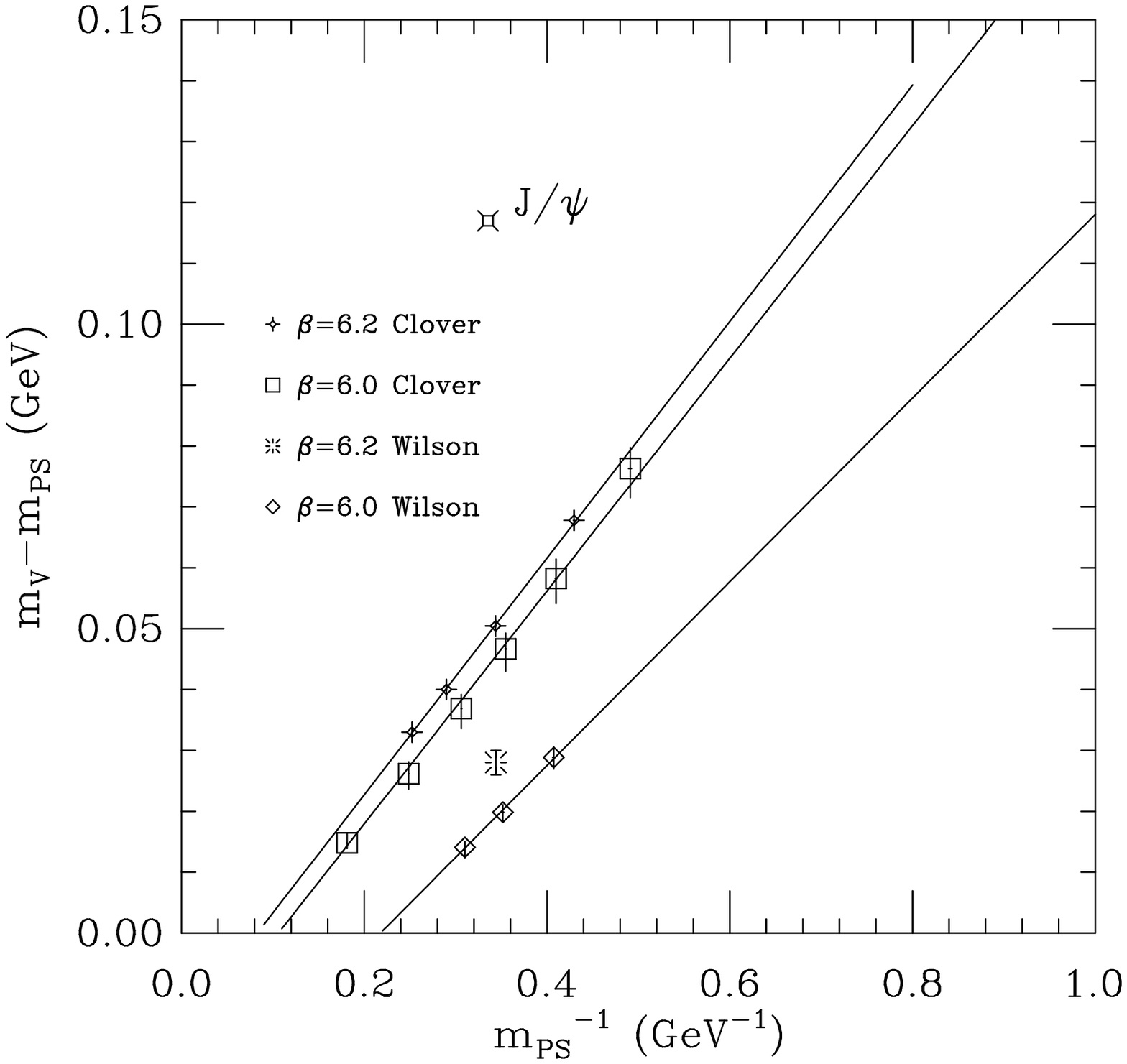}
}{saraQQhyp}

\fltfig{Heavy quark extrapolation of the quarkonium $\chi_{c1} - \chi_{c0}$ fine splitting
at $\beta=6.0$ using degenerate quarks and the $\gamma$ operators. This splitting is
clearly non-zero, but appears to be
significantly underestimated at this lattice spacing. Whether this
effect is due to quenching or discretisation errors cannot be understood
without corresponding simulations on other lattice spacings. The axes are scaled using the
string tension.
}{
\pspicture{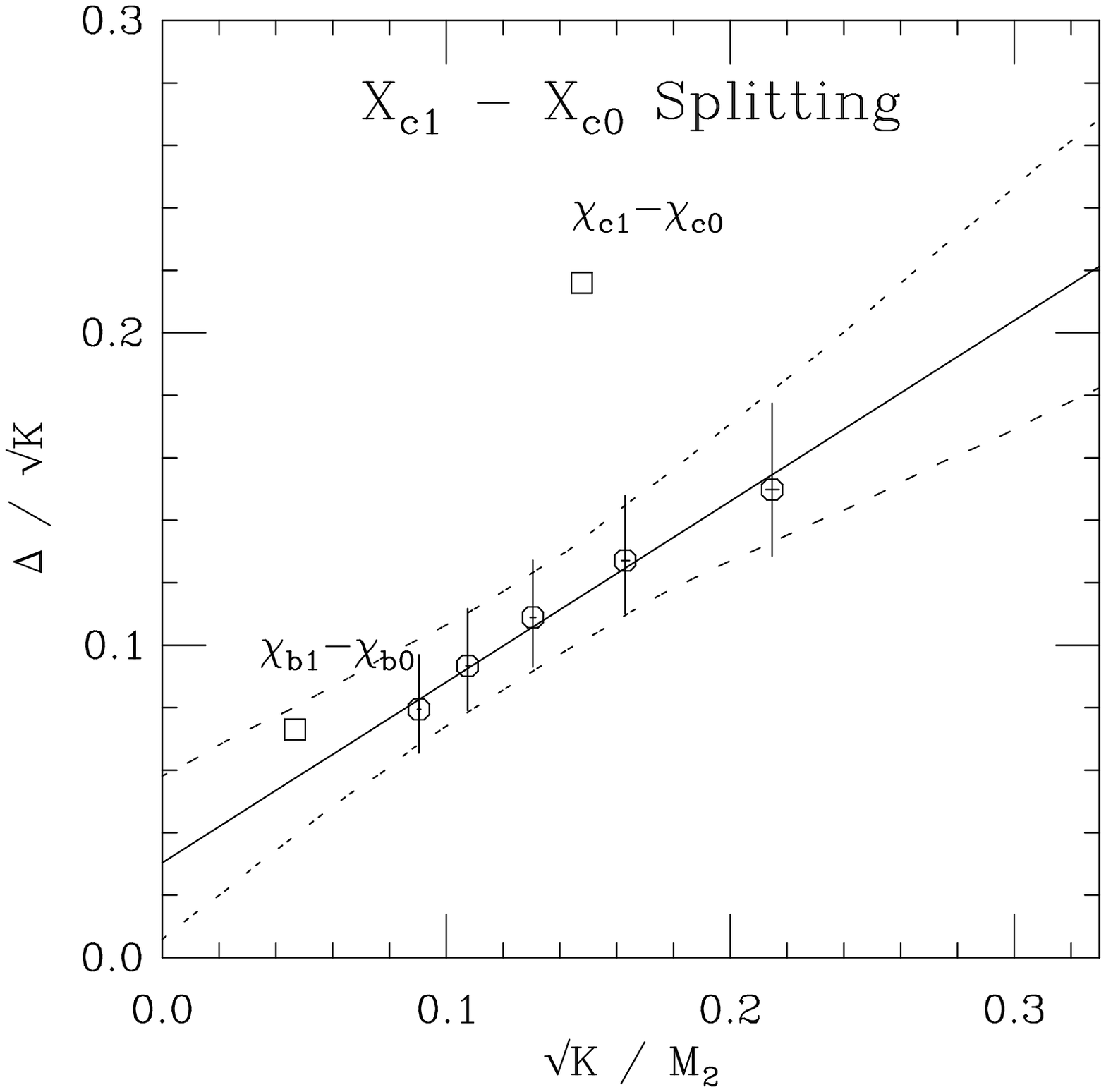}
}{FigHH_Xc0_Xc1fine60}

\fltfig{Heavy quark extrapolation of quarkonium $\chi_{c1} - h_c$ fine splitting
at $\beta=6.0$ using degenerate quarks and the $\gamma$ operators. This is statistically
consistent with zero at all masses simulated. The axes are scaled using the string tension.
}{
\pspicture{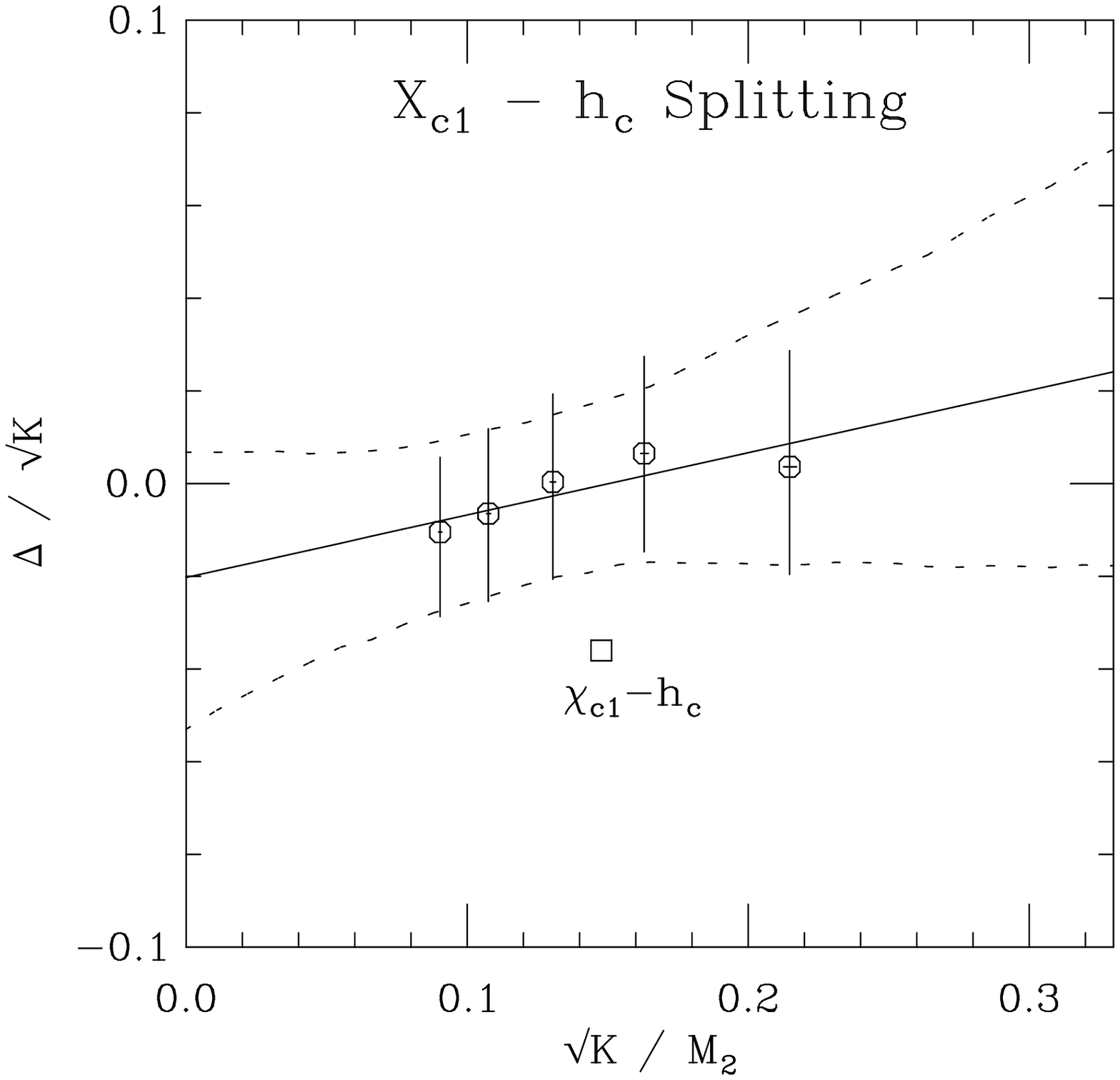}
}{FigHH_Xc1_Hcfine60}

\fltfig{Heavy quark extrapolation of quarkonium $\chi_{c2} - h_c$ fine splitting at $\beta=6.0$.
Here we use non-degenerate quarks $\kappa_1 \in \{0.114,0.118,0.122,0.126\}$ and
$\kappa_2 = 0.126$, and plot the difference between the mass of the $2^{++}$ state using
the differential operator and the $1^{++}$ using the $\gamma$ operator versus the inverse
kinetic mass of the corresponding pseudoscalar meson. The 
systematic errors in making predictions for degenerate quarkonia are expected to be 
below statistical errors for $c\bar{c}$, but are poorly controlled for
$b\bar{b}$. The results are scaled using the string tension.}{
\pspicture{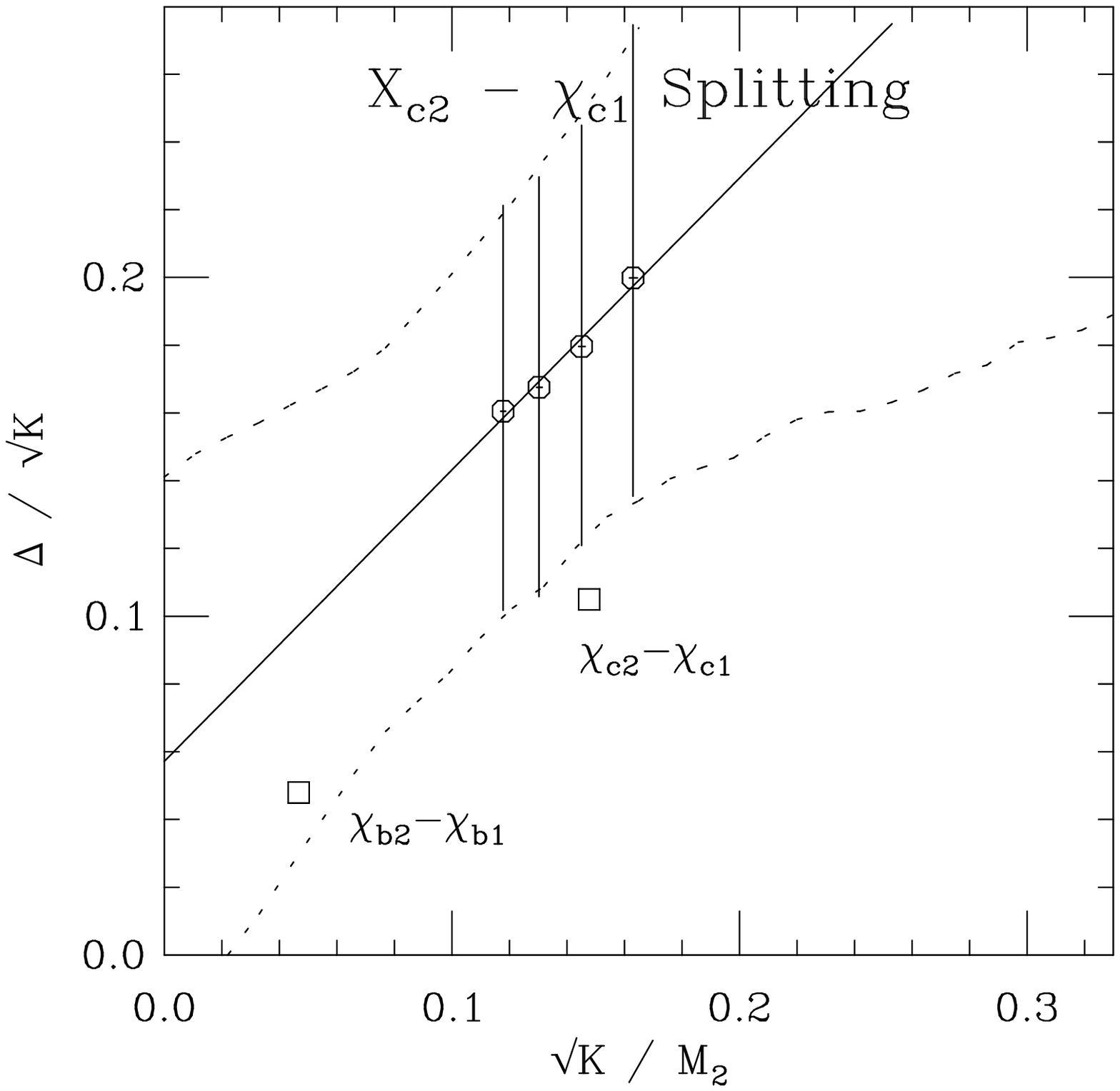}
}{FigHH_Xc2_Hcfine60}

\fltfig{Heavy quark extrapolation of quarkonium $\chi_{c2} - \chi_{c0}$ fine splitting
at $\beta=6.0$ in a similar manner to figure~\ref{FigHH_Xc2_Hcfine60}. The axes are scaled
using the string tension.
}{
\pspicture{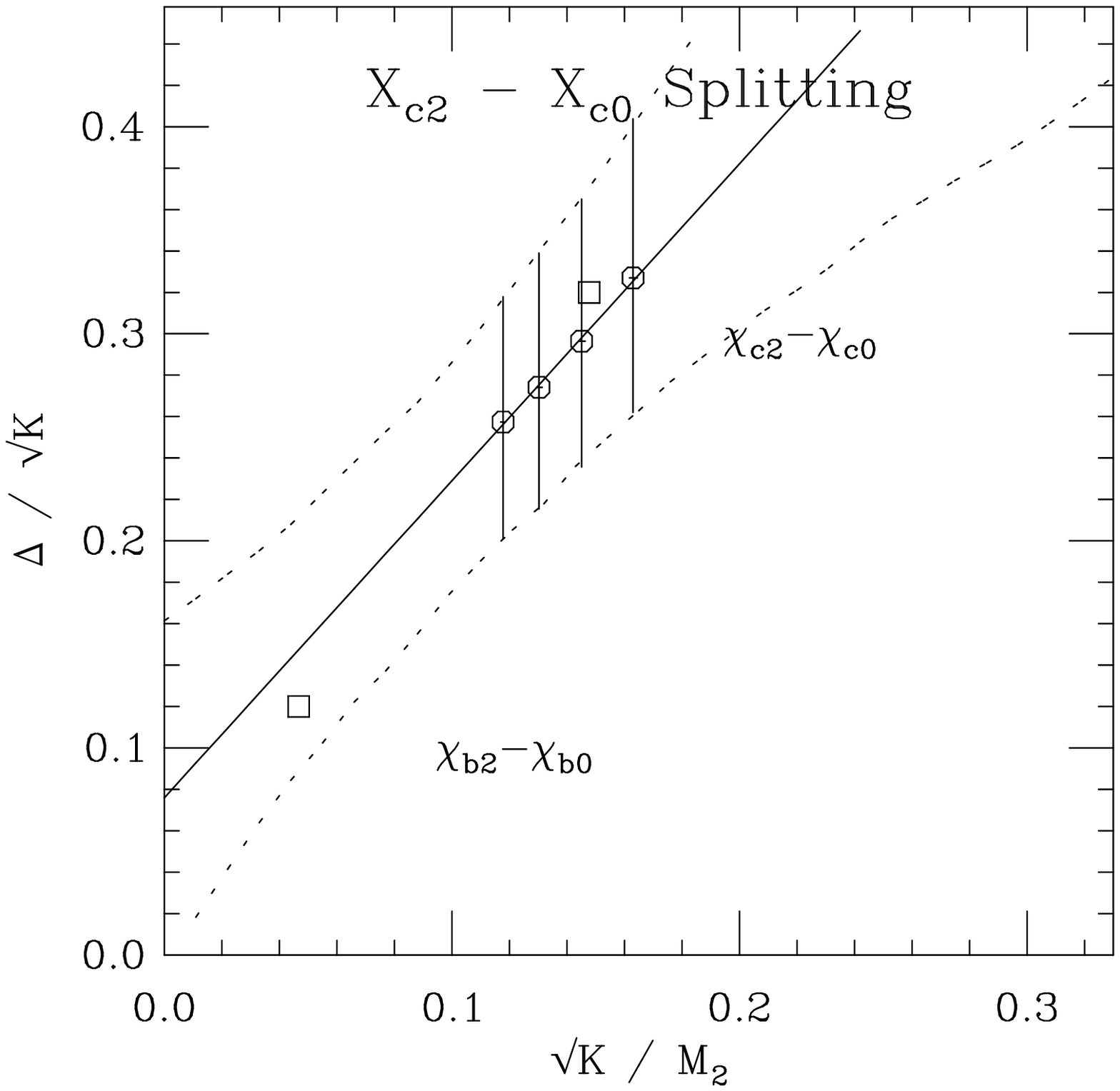}
}{FigHH_Xc2_Xc0fine60}

\fltfig{Charmonium spectrum diagram using the string tension to set the scale at $\beta=6.0$. 
Broad agreement is seen, but it appears that the spin structure is underestimated.
}{
\renewcommand{\yorg}{-2900}

\begin{minipage}[ht]{14cm}
\vspace{1in}
\setlength{\unitlength}{.0035in}
\begin{picture}(300,700)(-120,-10)
\spectrum{800}{3000}{3.0}{3500}{3.5}{3700}{GeV}

\state{\eta_c}{2980}{3015}{7}{7}{50}
\state{J/\psi}{3097}{3086}{2}{2}{200}
\state{\chi_{c0} }{3415}{3428}{15}{15}{430}
\state{\chi_{c1}}{3510}{3478}{14}{14}{550}
\state{\chi_{c2}}{3555}{3572}{26}{28}{670}
\state{h_{c}}{3526}{3475}{19}{19}{900}

\end{picture}
\end{minipage}

}{FigHH_Spectrum}

\fltfig{Upsilon spectrum diagram using string tension to set scale at $\beta=6.0$.
Clearly the $S-P$ splitting is underestimated when using the same scale as the
charmonium simulation. This effect is universally believed to be due to quenching.
It is commonly absorbed by using the S-P splitting to set the scale in a system
dependent manner, which would involve
a loss of predictive power here. Presenting our data in this manner shows that any single
quenched theory cannot simultaneously get both spectra correct.
}{
\renewcommand{\yorg}{-9300}
\begin{minipage}[ht]{14cm}
\vspace{1in}
\setlength{\unitlength}{.0035in}
\begin{picture}(300,700)(-120,-10)
\spectrum{1000}{9500}{9.5}{10000}{10.0}{10200}{GeV}
\state{\eta_b}{15000}{9434}{7}{7}{50}
\state{\Upsilon}{9460}{9455}{2}{2}{200}
\state{\chi_{b0} }{9860}{9776}{15}{15}{430}
\state{\chi_{b1}}{9892}{9800}{14}{14}{550}
\state{\chi_{b2}}{9913}{9854}{26}{28}{670}
\state{h_{b}}{15000}{9790}{19}{19}{900}
\end{picture}
\end{minipage}

}{FigHH_UpsSpectrum}


\fltfig{Scaling behaviour of $J/\psi$ and $\Upsilon$ hyperfine splitting between
$\beta=6.0$ and $\beta=6.2$. Results scale well with both the string tension
and $M_\rho$. We have no value for the $S-P$ splitting at $\beta=6.2$; it is possible
that one would obtain results with $S-P$ which scale, but yield
different continuum limit (particularly for $\Upsilon$), the discrepancy being a quenching error. }
{
\epsfaxhax{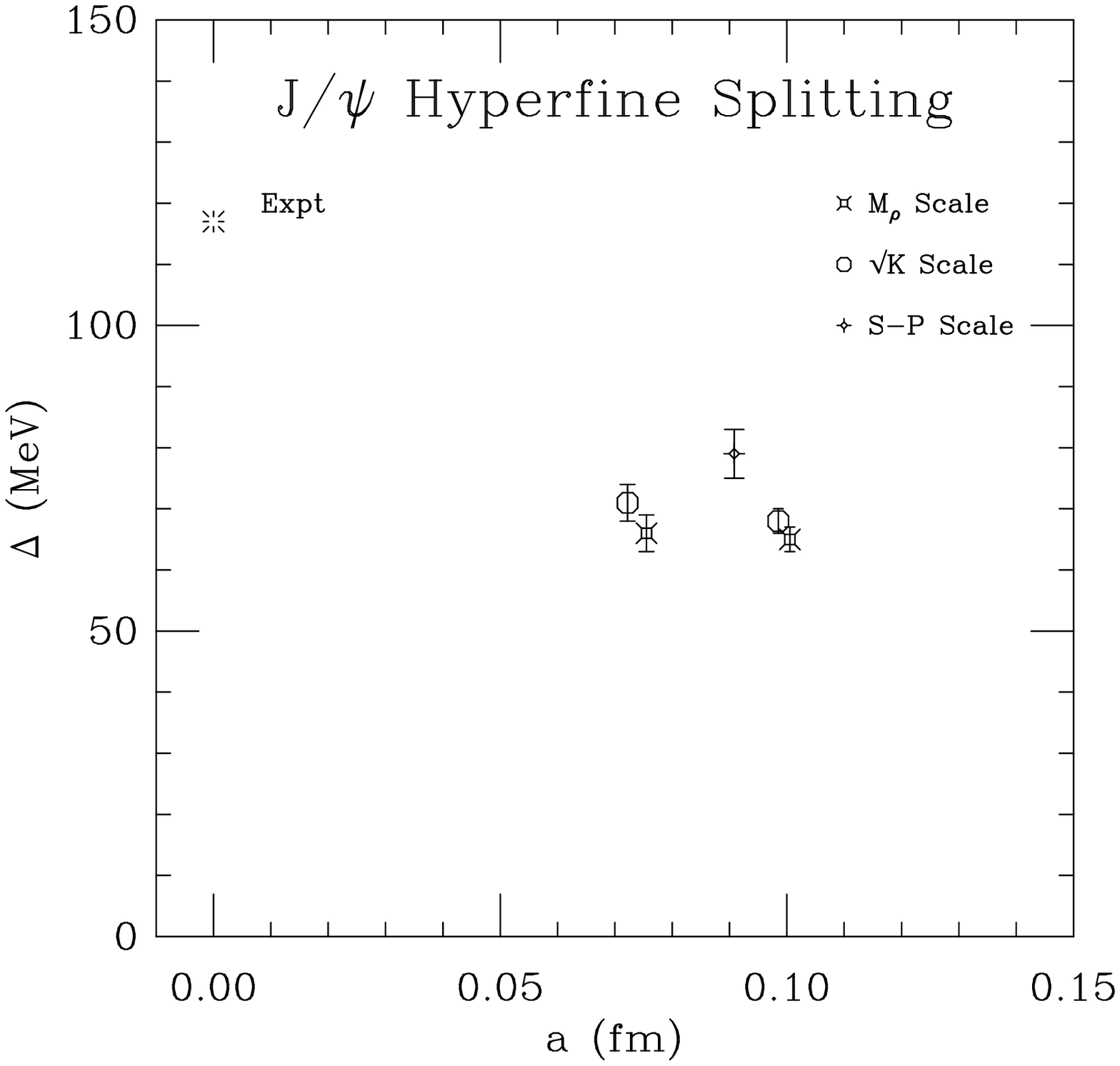}
{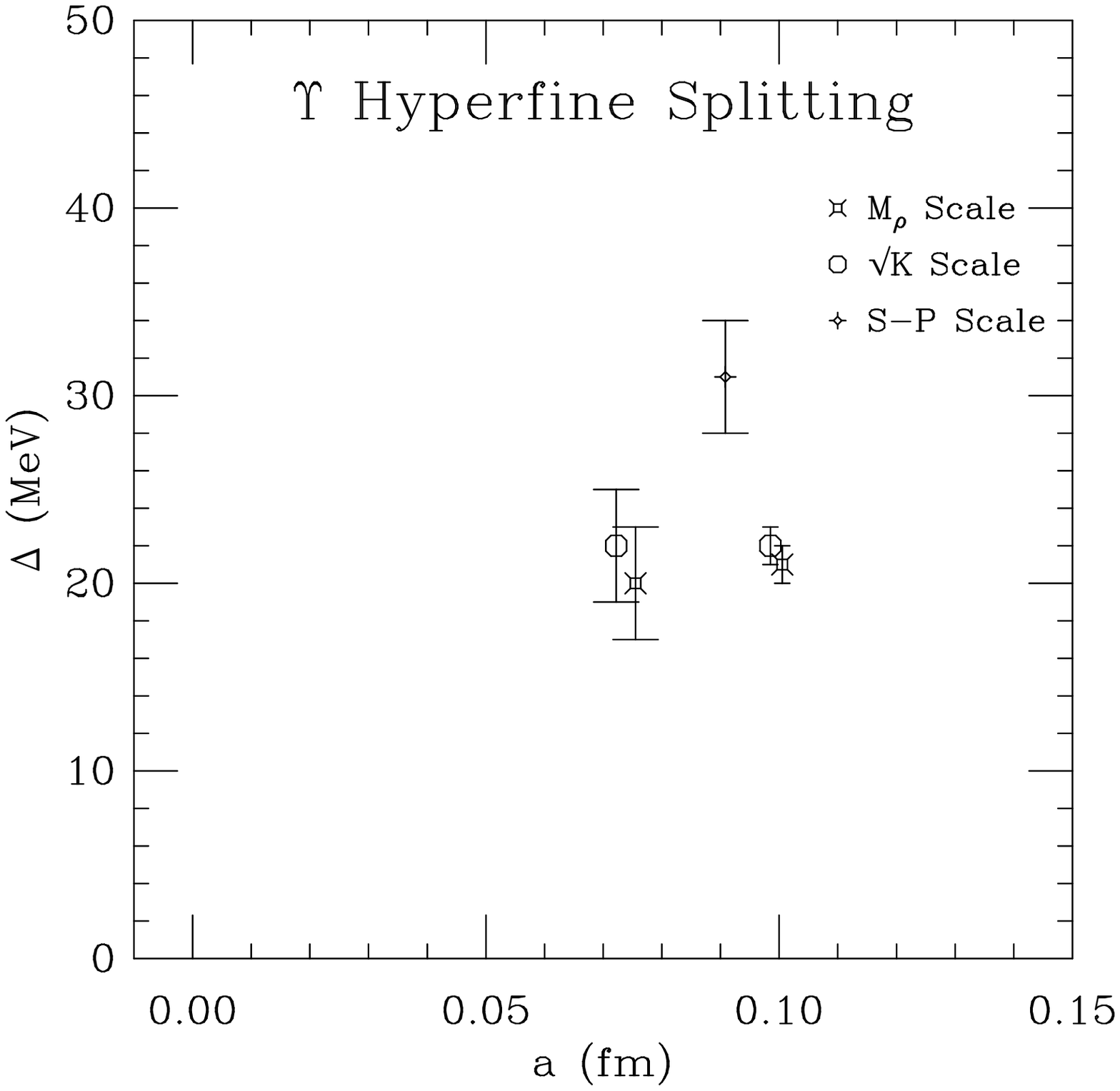}
}
{FigUpsHyperfineContinuum}

\fltfig{Scaling Behaviour of $m_{\rm charm}^{pole}$. The error here is
dominated by the uncertainty in $q^*$. It is plausible that the results
from the method 1 (diamonds) and method 2 (squares) will agree in the continuum limit.
}
{\pspicture{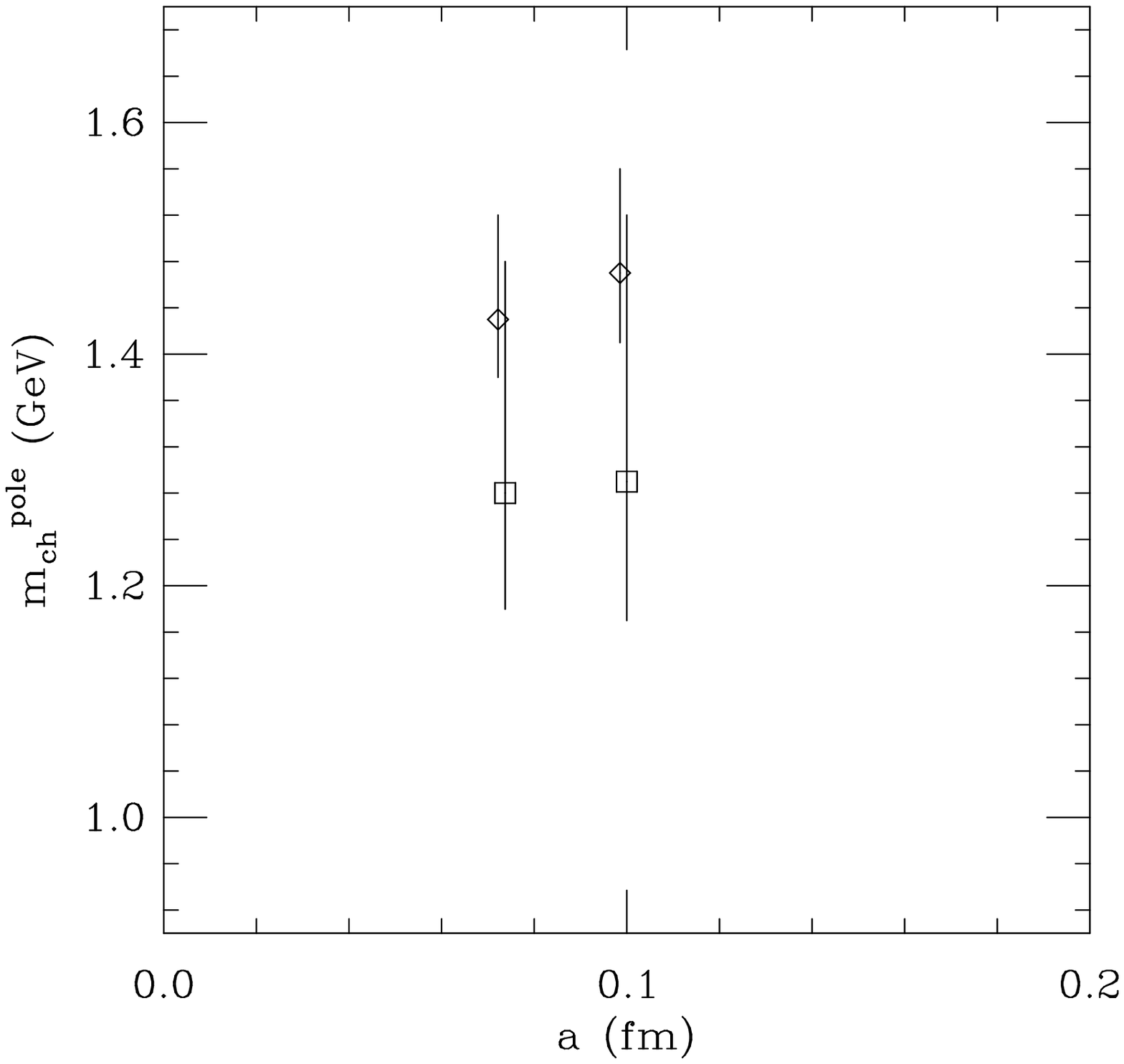}}
{FigCharmMass}

%
%

\flttab{$\beta=6.0$ heavy quark kappas}
{\begin{tabular}{|cccc|}
$\kappa$  & $aM_{PS}(Q\bar{Q})$&Fuzzing Radius& MR Iterations \\
\hline
0.13000& 0.9283(7)&3&95\\ 
0.12600& 1.1618(6)&3&75\\
0.12200& 1.3755(6)&3&65\\
0.11800& 1.5751(6)&3&60\\
0.11400& 1.7644(6)&3&55\\
\end{tabular}
}
{TabHeavy60Kappas}

\flttab{Meson operators}{
\begin{tabular}{|rr|c|}
State      & $J^{PC}$ & Operators \\
\hline
$^1S_0$    & $0^{-+}$ & $\bar{\psi} \gamma_5 \psi$           \\
$^3S_1$    & $1^{--}$ & $\bar{\psi} \gamma_i \psi$           \\
$^1P_1$    & $1^{+-}$ & $\bar{\psi} \sigma_{ij}\psi$ , $\bar{\psi} \gamma_5 \Delta_i \psi$ \\
$^3P_0$    & $0^{++}$ & $\bar{\psi} \psi$ , $\bar{\psi} \sum \gamma_i \Delta_i \psi$                  \\
$^3P_1$    & $1^{++}$ & $\bar{\psi} \gamma_i \gamma_5 \psi$ , $\bar{\psi} \{ \gamma_i \Delta_j - \gamma_j \Delta_j \} \psi$ \\
$^3P_2$    & $2^{++}$ & $\bar{\psi} \{\gamma_i  \Delta_i - \gamma_j \Delta_j \}\psi$ ~ {\rm E rep}\\ 
           &          & $\bar{\psi} \{\gamma_i  \Delta_j + \gamma_j \Delta_i \}\psi$  {\rm T rep}\\
\end{tabular}
}{tab:operators}

\flttab{$\beta=6.0$ fitted timeslice ranges}
{\begin{tabular}{|c|c|c|c|}
State & Pseudoscalar & Vector & P-states \\
\hline
Fit Range & 8-20 (FL,LL) & 8-20 (FL,LL) & 6-18 (FL,LL) \\
\end{tabular}
}{Tab60FitRanges}

\flttab{Heavy-heavy S states}{
\begin{tabular}{|c|c|c|c|c|}
State & $\chi^2/dof$ & Q & $M_{GS}$ & $M_{EX}$ \\
\hline
Pseudo (3000,3000)      & 0.83  & 0.68  & 0.9283(7)     & 1.41(3)\\
\hline
Pseudo (2600,2600)      & 0.88  & 0.61  & 1.1619(6)     & 1.60(2)\\
\hline
Pseudo (2200,2200)      & 0.99  & 0.47  & 1.3755(6)     & 1.79(2)\\
\hline
Pseudo (1800,1800)      & 1.2   & 0.25  & 1.5751(6)     & 1.97(1)\\
\hline
Pseudo (1400,1400)      & 1.4   & 0.093 & 1.7644(6)     & 2.14(1)\\
\hline
\hline
Vector (3000,3000)      & 1.0   & 0.44  & 0.9780(9)     & 1.49(2)\\
\hline
Vector (2600,2600)      & 0.95  & 0.52  & 1.1990(8)     & 1.66(2)\\
\hline
Vector (2200,2200)      & 1.0   & 0.44  & 1.4052(7)     & 1.83(1)\\
\hline
Vector (1800,1800)      & 1.2   & 0.26  & 1.5998(7)     & 2.00(1)\\
\hline
Vector (1400,1400)      & 1.4   & 0.12  & 1.7854(7)     & 2.17(1)
\end{tabular}
}{TabHHSstates}
\flttab{Heavy-heavy P states}{
\begin{tabular}{|c|c|c|c|c|}
State & $\chi^2/dof$ & Q & $M_{GS}$ & $M_{EX}$ \\
\hline
$0^{++}$ (3000,3000)  & 1.1   & 0.37  & 1.161(7)      & 1.76(10)\\
\hline
$0^{++}$ (2600,2600)  & 0.92  & 0.56  & 1.376(7)      & 1.89(7)\\
\hline
$0^{++}$ (2200,2200)  & 0.91  & 0.57  & 1.578(6)      & 2.04(5)\\
\hline
$0^{++}$ (1800,1800)  & 1.0   & 0.45  & 1.770(6)      & 2.19(4)\\
\hline
$0^{++}$ (1400,1400)  & 1.2   & 0.28  & 1.955(7)      & 2.34(4)\\
\hline
\hline
$1^{++}$ (3000,3000)  & 0.62  & 0.90  & 1.193(7)      & 1.76(7)\\
\hline
$1^{++}$ (2600,2600)  & 0.56  & 0.94  & 1.404(7)      & 1.91(6)\\
\hline
$1^{++}$ (2200,2200)  & 0.65  & 0.87  & 1.602(6)      & 2.07(5)\\
\hline
$1^{++}$ (1800,1800)  & 0.79  & 0.73  & 1.791(7)      & 2.22(4)\\
\hline
$1^{++}$ (1400,1400)  & 0.91  & 0.58  & 1.972(7)      & 2.37(4)\\
\hline
\hline
$1^{+-}$ (3000,3000)  & 0.56  & 0.94  & 1.193(8)      & 1.67(6)\\
\hline
$1^{+-}$ (2600,2600)  & 0.42  & 0.99  & 1.402(7)      & 1.86(5)\\
\hline
$1^{+-}$ (2200,2200)  & 0.51  & 0.96  & 1.602(7)      & 2.04(5)\\
\hline
$1^{+-}$ (1800,1800)  & 0.70  & 0.83  & 1.792(7)      & 2.21(4)\\
\hline
$1^{+-}$ (1400,1400)  & 0.86  & 0.64  & 1.975(7)      & 2.37(4)\\
\end{tabular}
}{TabHHPstates}

\flttab{Momenta or kinetic mass analyses}{
\begin{tabular}{|ccc|}
\multicolumn{3}{|c|}{Momentum Directions}\\
\hline
&(0,0,0)&\\
(1,0,0)&(0,1,0)&(0,0,1)\\
(1,1,0)&(1,0,1)&(0,1,1)\\
&(1,1,1)&\\
(2,0,0)&(0,2,0)&(0,0,2)\\
\end{tabular}
}{TabMom60}

\flttab{Heavy-heavy $M_2(PS)$}{
\begin{tabular}{|c|c|c|}
State & $\chi^2/dof$ & $M_{2}$ \\
\hline
Pseudo (3000,3000)      & 1.0   & 1.024(11)\\
\hline
Pseudo (2600,2600)      & 0.14  & 1.350(11)     \\
\hline
Pseudo (2200,2200)      & 0.14  & 1.686(12)     \\
\hline
Pseudo (1800,1800)      & 0.6   & 2.05(1)\\
\hline
Pseudo (1400,1400)      & 0.9   & 2.43(1)\\
\hline
Pseudo (2600,1400)      & 0.12  & 1.87(1)\\
\hline
Pseudo (2600,1800)      & 0.06  & 1.69(1)\\
\hline
Pseudo (2600,2200)      & 0.025 & 1.51(1)\\
\end{tabular}
}{TabHHM2}

\flttab{$\beta=6.0$ heavy-heavy mass splittings}
{
\begin{tabular}{|c|c|c|c|c|}
Splitting (MeV)& $\sqrt{K}$ & $M_\rho$ & $Q\bar{Q}$ S-P Splitting &Experiment\\
\hline
$J/\psi - \eta_c$	&68(2)&65(2)&79(4)&117\\
$^1P_1 - \bar{S}$	&418(13)&408(16)&-&458\\
$\chi_{c2} - \chi_{c1}$	& 81(28)&78(27)&93(33)&46\\
$\chi_{c1} - \chi_{c0}$	& 51(7) &49(7)&59(11)&95\\
$\chi_{c2} - \chi_{c0}$	&133(28)&128(28)&153(35)&141\\
\hline
$\Upsilon - \eta_b$	&22(1)&21(1)&31(3)& -\\
$^1P_1 - \bar{S}$	&366(17)&358(20)&-&460\\
$\chi_{b2} - \chi_{b1}$	&43(30)&42(30)&56(32)&21\\
$\chi_{b1} - \chi_{b0}$	&26(9)&25(9)&33(10)&32\\
$\chi_{b2} - \chi_{b0}$	&65(30)&63(27)&85(32)&53\\
\end{tabular}
}{Tab60Quarkonium}

\flttab{$\beta=6.2$ heavy-heavy mass splittings}
{
\begin{tabular}{|c|c|c|c|}
Splitting (MeV)& $\sqrt{K}$ & $M_\rho$ & Experiment\\
\hline
$J/\psi - \eta_c$& 72(3)&66(3)&117\\
\hline
$\Upsilon - \eta_b$     &22(3)&20(3)&-\\
\end{tabular}
}{Tab62Quarkonium}

\end{document}